\documentclass[pdflatex,sn-mathphys]{sn-jnl}
\usepackage{color,soul}

\usepackage{etoolbox}
\makeatletter
\patchcmd{\ps@headings}
{\hbox to \hsize{\hfill Springer Nature 2021 \LaTeX\ template\hfill}}
{\hbox to \hsize{\hfill \hfill}}
{}
{}
\patchcmd{\ps@titlepage}
{\hbox to \hsize{\hfill Springer Nature 2021 \LaTeX\ template\hfill}}
{\hbox to \hsize{\hfill \hfill}}
{}
{}
\makeatother

\jyear{2021}%

\theoremstyle{thmstyleone}%
\theoremstyle{thmstyletwo}%

\theoremstyle{thmstylethree}%

\begin{document}

\title[Article Title]{On-device Synaptic Memory Consolidation using Fowler-Nordheim Quantum-tunneling}

\author[1]{\fnm{Mustafizur} \sur{Rahman}}\email{mustafizur.rahman@wustl.edu}

\author[1]{\fnm{Subhankar} \sur{Bose}}\email{b.subhankar@wustl.edu}

\author*[1,2]{\fnm{Shantanu} \sur{Chakrabartty}}\email{shantanu@wustl.edu}

\affil[1]{\orgdiv{Department of Electrical and Systems Engineering}, \orgname{Washington University in St. Louis}, \orgaddress{\city{St. Louis}, \postcode{63130}, \state{MO}, \country{USA}}}
\affil[2]{\orgdiv{Department of Biomedical Engineering}, \orgname{Washington University in St. Louis}, \orgaddress{\city{St. Louis}, \postcode{63130}, \state{MO}, \country{USA}}}


\abstract{Synaptic memory consolidation has been heralded as one of the key mechanisms for supporting continual learning in neuromorphic Artificial Intelligence (AI) systems. Here we report that a Fowler-Nordheim (FN) quantum-tunneling device can implement synaptic memory consolidation similar to what can be achieved by algorithmic consolidation models like the cascade and the elastic weight consolidation (EWC) models. The proposed FN-synapse not only stores the synaptic weight but also stores the synapse's historical usage statistic on the device itself. We also show that the operation of the FN-synapse is near-optimal in terms of the synaptic life-time and we demonstrate that a network comprising FN-synapses outperforms a comparable EWC network for a small benchmark continual learning task. With an energy foot-print of femtojoules per synaptic update, we believe that the proposed FN-synapse provides an ultra-energy-efficient approch for implementing both synaptic memory consolidation and persistent learning.}




\maketitle
\begin{figure}[b]
\centering
\includegraphics[width=12cm]{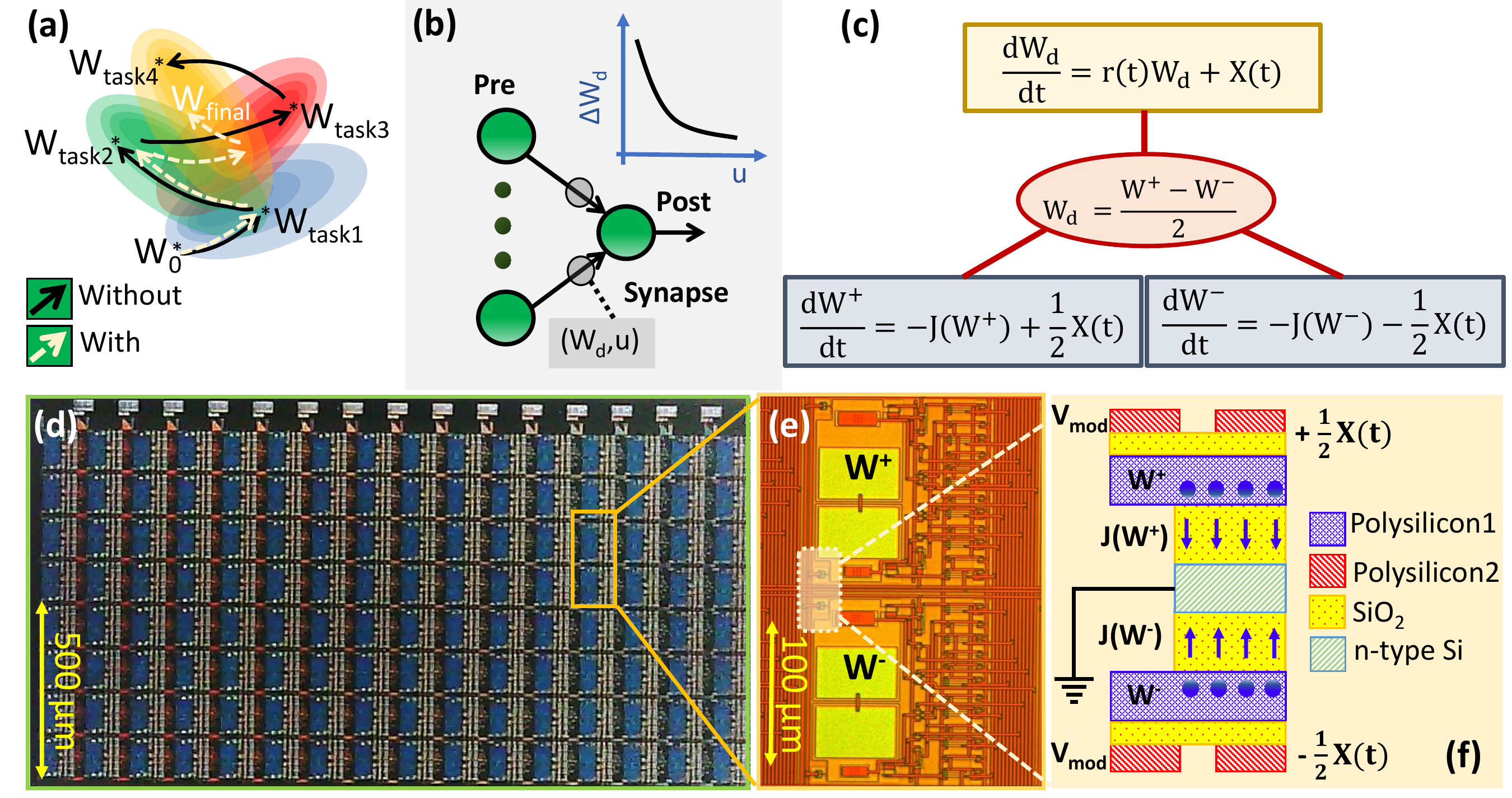}
\caption{Key concepts underlying the proposed FN-synapse device and synaptic memory consolidation: (a) Trajectory followed by synaptic weights during multi-task learning {\it with} and {\it without} memory consolidation; (b) A synaptic element supporting memory consolidation which stores both the magnitude $W_{d}$ of the weight and its usage $u$ where the propensity of successive weight updates $\Delta W_{d}$ changes with usage $u$; (c) Differential weight update dynamics in a single FN-synapse; (d) Micrograph of the fabricated FN-synapse array and (e) a single FN-synapse element; and (f) cross-sectional view of the FN-synaptic device showing the differential charge-storage elements ($W^+,W^-$), charge leakage elements ($J(W^+),J(W^-)$) and signal coupling elements ($\pm\frac{1}{2}X(t),V_{mod}$).}\label{Fig1}
\end{figure}

\section{Introduction}\label{intro}

\par There is a growing evidence from the field of neuroscience and neuroscience inspired AI about the importance of implementing synapses as a complex high-dimensional dynamical system~\cite{Ben15,FUSI05} as opposed to a simple and a static storage element, as depicted in standard neural networks~\cite{net}. This dynamical systems viewpoint has been motivated by {\it metaplasticity} observed in biological synapses~\cite{yan09,yan14} where the synaptic plasticity (or ease of update) can vary depending on age and task specific information that is accumulated during the process of learning. The process called {\it synaptic memory consolidation} is claimed to endow biological neural networks with an enhanced memory storage capacity and with the ability to learn multiple tasks over its lifetime without catastrophically forgetting old tasks~\cite{yan09}. In artificial neural networks (ANN) task specific synaptic memory consolidation has also been mimicked in the context of continual learning, as illustrated in Fig.~\ref{Fig1} (a). Without memory consolidation, a network learning a new task (or equivalently minimizing some objective function) will seek a solution that is optimal for the current task and in the process will forget the parameters learned during the previous tasks. On the other hand, a synaptic network with memory consolidation will learn the new task such that the new parameters don't significantly deviate from the parameters learned during the previous tasks. As a result, the network will seek solutions that work well for as many tasks as possible~\cite{ewc}. In literature several computational models have been proposed to achieve synaptic memory consolidation. Examples range from the complex cascade models~\cite{Ben15} to the elastic weight consolidation (EWC) models~\cite{ewc}. All these models follow a unified theme that each synapse should not only store the network weight but also an adaptive measure of the weight’s uncertainty or importance~\cite{ewc,oewc,rewc,Zenke,imm,mas}. This is illustrated in Fig.~\ref{Fig1} (b) where the synapses are shown to store a usage/age parameter $u$ in addition to the usual synaptic weight $W_d$. The dynamics of the synaptic weight update $\Delta W_d$ is then governed by $u$ which can take different temporal profiles and hence can implement different consolidation models. One such profile is illustrated in Fig.~\ref{Fig1} (b) where the synapses are shown to become more rigid (less plastic) with usage. However, mathematically the different synaptic consolidation approaches can be generalized using a unified dynamical systems model

\begin{equation}\label{eq01}
\begin{split}
\frac{dW_{d}}{dt}=r(t) W_{d} + X(t)\\
\end{split}
\end{equation}
which describes the propensity to change the stored weight $W_{d}$ in response to an external input $X(t)$ applied at a time instant $t$. Note that the magnitude of weight change in equation~\ref{eq01} is proportional to previous value of $W_{d}$. $r(t)$ is the time varying function that models the dynamics of the synaptic plasticity as a function of the history of synaptic activity (or its usage). For cascade models, equation~\ref{eq01} is implemented as a weighted linear superposition of the inputs $X(t)$~\cite{Ben15}, whereas in an EWC model $r(t)$ is a function of the Fisher information~\cite{ewc}. For both models it has been shown that when the synaptic network is subjected to random, uncorrelated memory patterns the strength of a ``memory" (typically defined in terms of signal-to-noise ratio) decays at an optimal rate of $1/\sqrt{t}$ over the effective synaptic life-time. However, the key difference between the EWC and the cascade models is in the manner in which the strength of a ``memory" degrades as the network approaches or exceeds its capacity. Because the cascade models incorporate a mechanism for ``forgetting" old patterns, the synaptic network is still able to learn and store new patterns, as a result, the network exhibits a gradual degradation in memory retention. In comparison, EWC network consolidates older memories over time and as the network approaches capacity, no new memory patterns can be stored. Once the network exceeds its capacity, EWC networks are known to suffer from {\it blackout catastrophe}~\cite{ewc,hop} where even the old memories are rapidly forgotten. However, when operating within the capacity limits, both types of consolidation models facilitate continuous learning of new patterns over long periods of time. From an implementation point-of-view both these consolidation models are algorithmic in nature and to-date no synaptic device has been reported that can mimic or combine the characteristics of both the cascade and the EWC models. In~\cite{Ben15}, a physical implementation of the cascade model has been suggested but requires dynamic coupling of multiple physical processes and state variables, and hence is not scalable for large-scale hardware implementation.

\par In this paper we present a physically realizable synaptic device that can demonstrate synaptic memory consolidation similar to the algorithmic cascade and the EWC models. The synaptic device implements the model in equation~\ref{eq01} using two differential non-linear dynamical systems, as shown in Fig.~\ref{Fig1} (c). $J(.)$ in Fig.~\ref{Fig1} (c) represents an arbitrary non-linear function and the synaptic weight $W_{d}$ is stored as a differential quantity $W_{d} = \frac{1}{2}(W^+ - W^-)$. $W_{d}$ is updated according to a differential input signal $\pm\frac{1}{2}X(t)$. In the Methods section~\ref{metwup} we show that under the assumption that $W_{d} << W_{c}$ where $W_c = \frac{1}{2}(W^+ + W^-)$ is a common-mode quantity, the update dynamics of the synapse can be expressed as

\begin{equation}\label{sec1_eq2}
\frac{dW_d}{dt}=-\left[\frac{d^{2}W_c}{dt^{2}}\left(\frac{dW_c}{dt}\right)^{-1}\right]W_d+X(t).
\end{equation}
Equation~\ref{sec1_eq2} shows that the weight update does not directly depend on the non-linear function $J(.)$ but implicitly through the common-mode $W_c$. Comparing equations~\ref{sec1_eq2} and~\ref{eq01} and relating to the previously reported time-evolution profile $r(t) \approx \mathcal{O}(1/t)$~\cite{ewc} (for random, uncorrelated memory patterns), equation~\ref{sec1_eq2} can be used to derive an optimal form of $J(.)$. In the Methods Section~\ref{metopup} we show that if the objective is to maximize the operational lifetime of the synapse, the optimal $J(.)$ takes the form $J(V) \propto V^2 \exp\left(-\beta/V\right)$, $\beta$ being a constant. The expression for $J(V)$ matches the expression for a Fowler-Nordheim (FN) quantum-tunneling current~\cite{lenz} indicating that optimal synaptic memory consolidation could be achieved by a physical device comprising of FN quantum-tunneling junctions.

Fig.~\ref{Fig1} (d)-(e) shows the micrograph of an synaptic array comprising of differential FN tunneling junctions fabricated in a standard silicon process. In the Methods Section~\ref{metsnr} we show the mapping of the dynamical systems model in Fig.~\ref{Fig1} (c) using the physical variables associated with FN quantum tunneling. Similar to our previous works~\cite{Zho17a,Zho19,secure}, the tunneling junctions have been implemented using polysilicon, Silicon-di-oxide and n-well layers as illustrated in Fig.~\ref{Fig1} (f). The polysilicon layer forms a floating-gate where the initial charge can be programmed using a combination of hot-electron injection or quantum-tunneling~\cite{adapt,Meh20}.The synaptic weight is stored as a differential voltage $W_d = \frac{1}{2}(W^+ - W^-)$ across two floating-gates as shown in Fig.~\ref{Fig1} (f). The voltages on the floating-gates $W^+$ and $W^-$ at any instant of time is modified by the differential signals $\pm \frac{1}{2}X(t)$ that are coupled onto the floating-gate. The dynamics for updating $W^+$ and $W^-$ are determined by the respective tunneling currents $J(.)$ which discharge the floating-gates. In the Supplementary Section I we describe the complete equivalent circuit along with the read-out mechanism used in this work for measuring $W_d$ of the FN-synapse shown in Fig~\ref{Fig1} (e). The presence of additional coupling capacitors in Fig~\ref{Fig1} (f) allows the plasticity of all the FN-synapses in an array to be adjusted using an external signal $V_{mod}$. Using this feature we show that the memory consolidation characteristics of the FN-synapse array can be adjusted to mimic either the cascade models (with different degrees of complexities) or the task-specific synaptic consolidation corresponding to the EWC model. Using a benchmark continual learning task we also show that a network of FN-synapses can remember older tasks better than a comparable EWC network or a network that uses a stochastic gradient descent (SGD) based learning.

\section{Results}\label{sec2}
\begin{figure}[b]
\centering
\includegraphics[width=12cm]{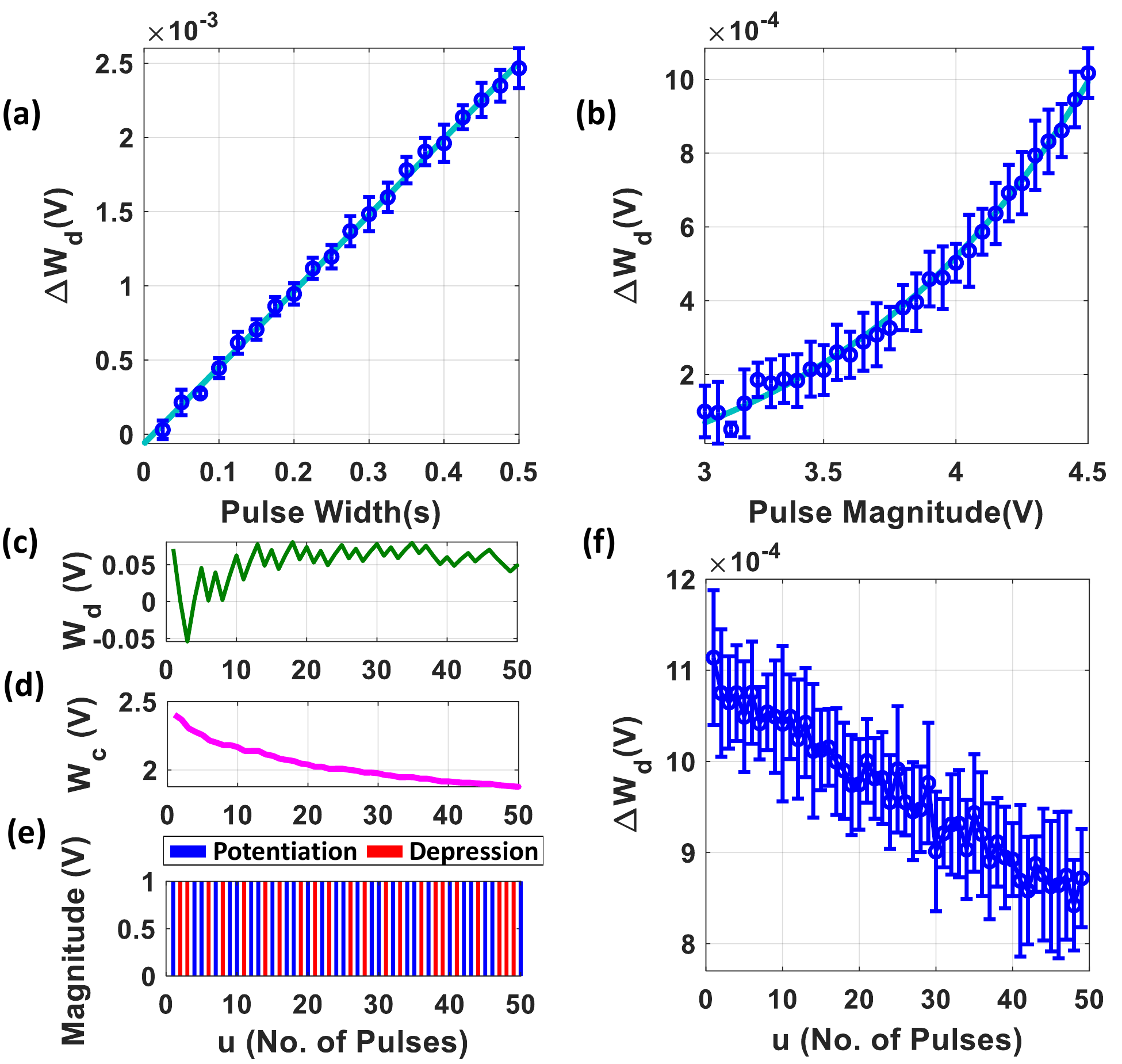}
\caption{Characterization of a single FN-synapse:(a) Dependence of change in magnitude of weight with change in pulse-width which follows a linear trajectory defined by $y=mx+c$ (where $m=0.005136$ and $c=-6.227\times 10^{-5}$). (b) Dependence on pulse magnitude of the input pulse which follows an exponential trajectory defined by $y=c\times exp(ax+b) + d$ (where $a=1$, $b=-6.611$, $c=0.009959$ and $d= -0.0002142$). (c) Bidirectional evolution of weight ($W_d$) and (d) trajectory followed by the common-mode tunneling node ($W_c$) when the synapse is subjected to (e) a random set of {\it potentiation} and {\it depression} pulses of equal magnitude and duration. (f) Change in the magnitude of successive weight updates ($\Delta W_d$) corresponding to repeated stimulus.}\label{Fig2}
\end{figure}
\subsection{FN-synapse Characterization}\label{21}

The first set of experiments were designed to understand how FN-synapses adapt in response to an external stimulation and store the synaptic weight and a measure of synaptic usage. Techniques to initialize the charge stored on the floating-gates in an FN-synapse can be found in the
Methods section~\ref{metprog} and in the Appendix. The tunneling barrier thickness in FN-synapse prototype shown in Fig.~\ref{Fig1} (c)-(d) was chosen to be greater than 12nm which makes the probability of direct tunneling of electrons across the barrier to be negligible. Also, when the electric potential of the tunneling nodes $W^+$ and $W^-$ are set to be less than 5V, probability of FN tunneling of electrons across the barrier becomes negligible. In this state, the FN-synapse behaves as a non-volatile memory storing a weight proportional to $W_{d}=W^+-W^-$. To increase the magnitude of the stored weight a differential input pulse $\pm \frac{1}{2}X$ is applied across the coupling capacitors. The electric potential of the floating-gate $W^-$ is pushed to a regime $(>7.5V)$ where the FN tunneling current $J(W^-)$ is now significant. At the same time the electric potential of the floating-gate $W^+$ is also pushed higher but with FN tunneling current $J(W^+) < J(W^-)$.  As a result the $W^-$ node discharges at a rate that is faster than the $W^+$ node. After the cessation of the input pulse, the potential of both $W^-$ and $W^+$ become less than 5V and hence the FN-synapse returns to its non-volatile state. Fig.~\ref{Fig2} (a) and (b) show the measured weight update $\Delta W_{d}$ in response to different magnitudes and duration of the input pulses. For this experiment the common-mode $W_{c}=\frac{1}{2}(W^++W^-)$ is held fixed. In Fig.~\ref{Fig2} (a) we can observe that $\Delta W_{d}$ changes linearly with pulse width for a fixed magnitude of input voltage pulses (= 4V). Fig.~\ref{Fig2} (b) shows that the updated $\Delta W_{d}$ changes exponentially with respect to the magnitude of the input pulses (duration = 100ms). The results show that pulse width modulation or pulse density modulation provides a more accurate control over the synaptic updates. When the common-mode $W_{c}$ is not held fixed, irrespective of whether the weight $W_d$ is increased or decreased (depending on the polarity of the input signal) the common-mode always decreases. Thus, $W_c$ could serve as an indicator of the usage of the synapse. The measured result in Fig.~\ref{Fig2} (c)-(e) show that an FN-synapse can store both the weight and the usage history. For this experiment we applied a series of {\it potentiation} and {\it depression} pulses of equal magnitude and duration to the FN-synapse, as shown in Fig.~\ref{Fig2} (e). Fig.~\ref{Fig2} (c) depicts the weight stored and how it evolves bidirectionally (like a random walk) due to the input pulses. Meanwhile, the common-mode potential $W_c$ decreases monotonically with the number of inputs irrespective of the polarity of the input, as shown in Fig.~\ref{Fig2} (d). Therefore, $W_c$ reliably tracks the usage history of the FN-synapse. Fig.~\ref{Fig2} (f) shows the {\it metaplasticity} exhibited by an FN-synapse where we measured $\Delta W_{d}$ as a function of usage $u$ by applying successive {\it potentiation} input pulses of constant magnitude (4V) and width (100ms). Fig.~\ref{Fig2} (f) shows that when the synapse is modulated with same excitation successively, the amount of weight update decreases monotonically with increasing usage, similar to the response illustrated in Fig.~\ref{Fig1} (b). 



\begin{figure}[b]
\centering
\includegraphics[width=12cm]{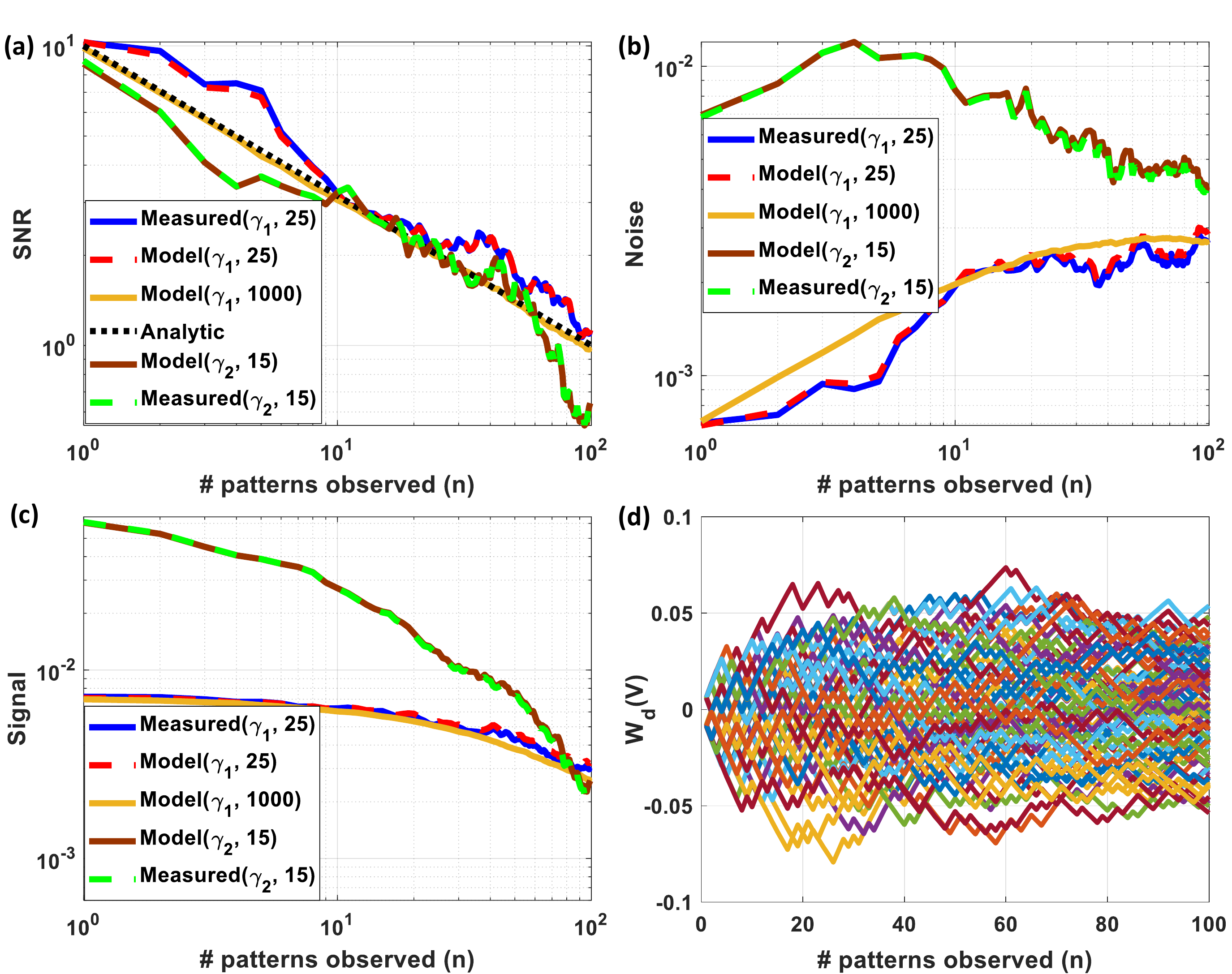}
\caption{Memory-Lifetime experiments: Comparison of (a) SNR, (b) noise strength and (c) signal strength for a network size of 100 synapse measured using the FN-Synapse chipset over 25 (for $\gamma_1$) and 15 (for $\gamma_2$) Monte-Carlo runs. The legends associated with the plots are specified as ($\gamma$, Number of Monte-Carlo runs). The numerical results obtained using the FN-synapse model (25 and 1000 Monte Carlo simulations for $\gamma_1$ and 15 simulations for $\gamma_2$). Here $\gamma_1>\gamma_2$ (d) Measured evolution of weights of the FN-synapse network. Each color corresponds to the weight of an individual FN-synapse in the network.}\label{Fig3}
\end{figure}

\subsection{FN-synapse Network Capacity and Memory lifetime}\label{subsec2}
The next set of experiments were designed to understand the memory consolidation characteristics for an FN-synapse array that is excited using a random binary input pattern ($potentiation$ or $depression$ pulses). This type of benchmark experiment is used extensively in memory consolidation
studies~\cite{Ben15,ewc} to compare the experimental results with analytical/theoretical results. A network comprising of $N$ FN-synapses is first initialized to store zero weights (or equivalently $W^- = W^+$). New memories were presented as random binary patterns ($N$ dimensional randam binary vector) that are applied to the $N$ FN-synapses through either $potentiation$ or $depression$ pulses. Each synaptic element was provided with balanced input i.e. equal number of $potentiation$ and $depression$ pulses. The goal of this experiment is to track the strength of a memory that is imprinted on this array in the presence of repeated new memory patterns. It can be envisioned that as additional new patterns are written to the same array, the strength of a specific memory will degrade. Similar to the previous studies~\cite{Ben15,ewc} we quantify this degradation in terms of signal-to-noise ratio (SNR). If $n$ denotes the number of new memory patterns that have been applied to the FN-synapse array, then the Methods section \ref{metsnr} shows that the retrieval memory signal $S(n)$ power, the noise $\nu(n)$ power and the $SNR(n)$ can be expressed analytically as
 
\begin{equation}\label{eq1}
\begin{split}
S^2(n)=\frac{1}{\left(n+\gamma\right)^2}\\
\nu^2(n)=\frac{n}{N(n+\gamma)^{2}}\\
SNR(n)=\sqrt{\frac{N}{n}}.
\end{split}
\end{equation}
Here $\gamma$ is a device parameter that depends on the initialization condition, material properties and duration of the input stimuli. Equation \ref{eq1} shows that the initial SNR is $\sqrt{N}$ and the SNR falls off according to a power-law decay with a slope of $\frac{1}{\sqrt{n}}$. Like previous consolidation studies~\cite{Ben15} we will assume that a specific memory pattern is retained as long as its SNR exceeds a predetermined threshold. Therefore, according to equations~\ref{eq1} the network capacity and memory lifetime for FN-synapse scales linearly with the size of the network $N$. These quantities for FN-synapse that defines its memory performance matches with the characteristic for a network of synapse trained using EWC. We verified the analytical expressions in equation \ref{eq1} for a network size of $N=100$ using results measured from the FN-synapse chipset. Details of the hardware experiment is provided in the Method Section~\ref{metsnrhard}. Fig.~\ref{Fig3} (a), (b), and (c) shows the SNR, noise and the retrieval signal obtained from the fabricated FN-synapse network for two different values of $\gamma$. We observe that the SNR obtained from the hardware results conform to the analytical expressions relatively well. The slight differences can be attributed to the Monte-Carlo simulation artifacts (only 25 and 15 iterations were carried out). In the Appendix we show verification of these analytic expressions using a software model of the FN-synapse. Details on the derivation of FN-synapse model is provided in the Methods Section~\ref{metsnrhard}. The simulated results in Fig.~\ref{Fig3} (a), (b) and (c) verifies that results from the software model can accurately track the hardware FN-synapse measurements for both values of $\gamma$ when subjected to the same stimuli. Therefore, FN-synapse and its software model can be used interchangeably. The results also show that when the number of iterations on the Monte-Carlo simulation is increased (1000 iterations), the simulated SNR closely approximates the analytic expression. This verifies that hardware FN-synapse is also capable of exactly matching the optimal analytic consolidation characteristics. 
Fig.~\ref{Fig3} (d) shows the measured evolution of each weight stored in the FN-synapse network where initially the weights grow quickly but after a certain number of updates settle to a steady value irrespective of new updates. This implies that the synapses have become rigid with an increase in its usage. This type of memory consolidation is also observed in EWC models which has been used for continual learning. However, note that unlike EWC models that need to store and update some measure of Fisher information, here the physics of the FN-synapse device itself can achieve EWC-like memory consolidation without any additional computation.

\subsection{Cascade FN-Synapse Models}\label{subseccas}    

\begin{figure}[b]
\centering
\includegraphics[width=6cm]{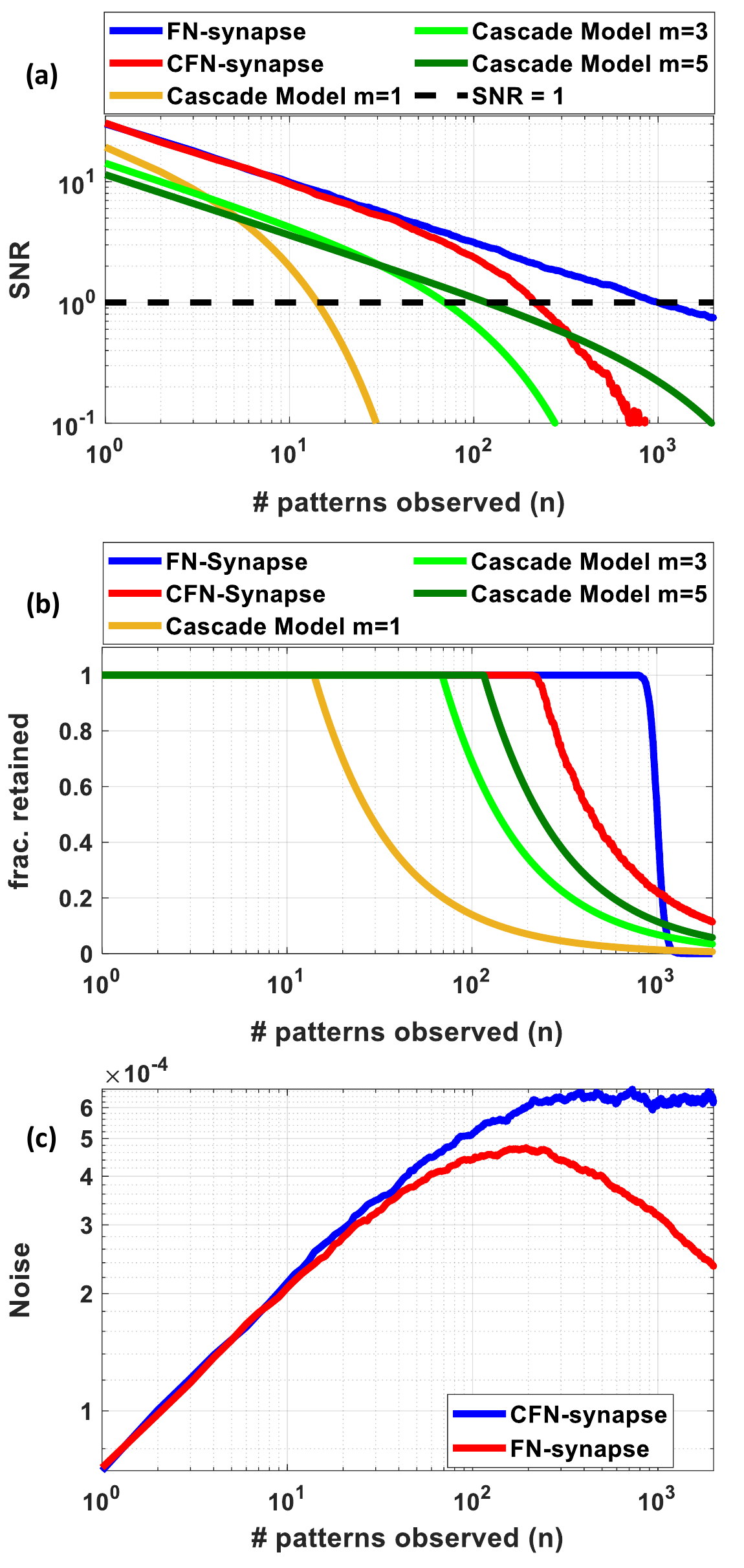}
\caption{Network capacity and saturation experiments: Comparison of (a) SNR, (b) fraction of patterns retained and (c) noise of networks composed of 1000 synapses following different synaptic models when exposed to 2000 patterns. CFN-synapse is the FN-synapse configured to mimic the cascade model.}\label{Fig4}
\end{figure}
In our next set of experiments we verify that the plasticity of FN-synapses can be adjusted to mimic the consolidation properties of both EWC and cascade models. While EWC models only allows for retention of old memories, cascade models allow for both memory retention and forgetting. As a result, cascade models avoid {\it blackout catastrophe} whereas an EWC network is unable to retrieve any previous memories or store new experiences as the network approaches its capacity. Cascade models allow the network to gracefully forget old memories and continue to remember new experiences indefinitely. For an FN-synapse network, a coupling capacitor in each synapse (shown in Fig.~\ref{Fig1} (f)) which is driven by a global signal $V_{mod}$ can control the plasticity of the FN-synapse and mimic the cascade model. Details of the modified cascaded FN-synapse (CFN-synapse) is provided in the Methods Section~\ref{metcfn}. To understand and compare the blackout catastrophe in FN-synapse models with cascade model we define the metric $frac. retained$ as the fraction of patterns whose SNR exceeds $1$. The SNR and fraction retained for CFN-synapse network of size $N=1000$ is shown in Fig. \ref{Fig4} (a) and (b) respectively together with those for cascade models of different levels of complexity~\cite{Ben15} (denoted by $m=1,..,5$) and FN-synapse. We can observe in Fig. \ref{Fig4} (a) that the FN-synapse network outperforms every other model in terms of SNR (similar to an EWC network). However, once the network capacity is reached (around 1000 patterns), FN-synapse forgets all observed patterns in addition to not forming any new memories as $frac. retained$ goes to zero. Whereas in the case for CFN-synapse the $frac. retained$ degrades slowly similar to that of the cascade models. This comes at a cost of network capacity which does not scale linearly with $N$ anymore. Nevertheless, we find that CFN-synapse is still able to match cascade models in regard to memory retention time while outperforming in initial SNR strength. In addition we have also verified that the synaptic strength of CFN-synapse is bounded similar to the cascade models. This can be observed in Fig.~\ref{Fig4} (c) which shows that the variance in retrieval signal (Noise) of both FN-synapse and CFN-synapse network remain bounded. Therefore, our synaptic models can exhibit responses similar to both EWC and cascade models while being physically realizable and scalable for large networks.

\subsection{Continual Learning using FN-synapse}\label{subseclearn}

The next set or experiments were designed to evaluate the performance of FN-synapse neural network for a benchmark continual learning task. A fully-connected neural network with two hidden layers was trained sequentially on multiple supervised learning tasks. Details of the neural network architecture and training are given in Methods Section~\ref{metnn} and Appendix Section \ref{asec5}. The network was trained on each task for a fixed number of epochs and after the completion of its training on a particular task $t_{n}$, the dataset from $t_{n}$ was not used for the succesive task $t_{n+1}$. 

The aforementioned tasks were constructed from the Mixed National Institute of Standards and Technology (MNIST) dataset, to adress the problem of classifying handwritten digits in accordance with schemes popularly used in several continual-learning literature \cite{NIPS18}. Also known as incremental domain learning using split-MNIST dataset, each task of this continual learning benchmark dictates the neural network to be trained as binary classifier which distinguishes between a set of two hand-written digits, i.e. the network is first trained to distinguish between the set $[0,1]$ as $t_{1}$ and is then trained to distinguish between $[2,3]$ in $t_{2}$, $[4,5]$ in $t_{3}$, $[6,7]$ in $t_{4}$ and $[8,9]$ in $t_{5}$. Thus, the network acts as an even-odd number classifier during every task.

\begin{figure}[b]
\centering
\includegraphics[width=12cm]{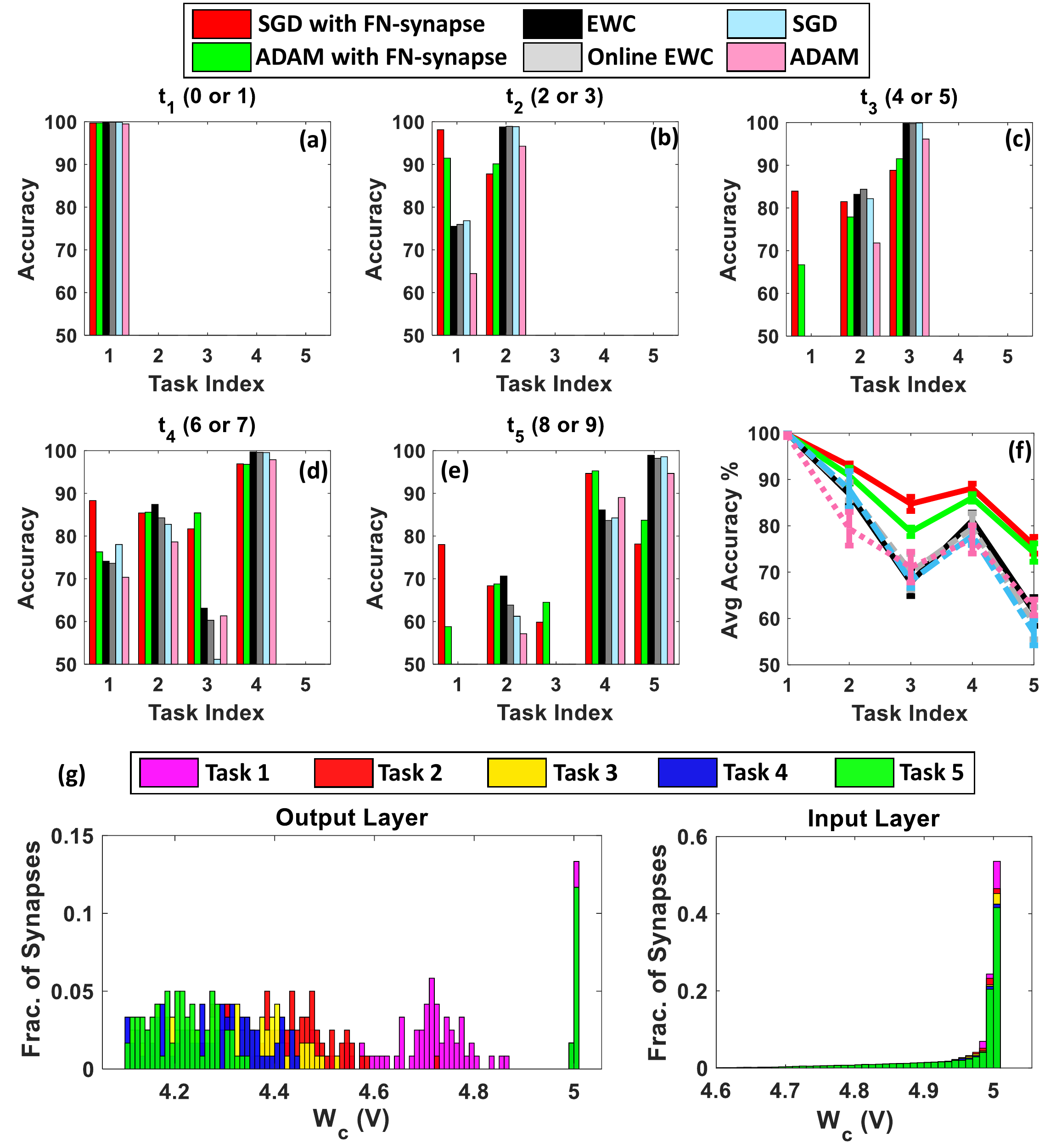}
\caption{Continual learning benchmarks results and insights: (a)-(e) Task-wise accuracy and (f) overall average accuracy comparison of SGD and ADAM with FN-synapse, ADAM with EWC and Online EWC, SGD and ADAM with conventional memory. (g) Distribution of the usage profile of weights in the output layer and the input layer of the FN-synapse neural network.}\label{Fig5}
\end{figure}
Fig. \ref{Fig5} (a)-(e) compares the task-wise accuracy of networks trained with different learning and consolidation approaches. Note here that the absence of a data-point corresponding to a particular approach indicates that the accuracy obtained is below $50\%$. All the approaches taken into consideration perform equally well at learning $t_{1}$ as illustrated in \ref{Fig5} (a). However, as the networks learn $t_{2}$ (see Fig. \ref{Fig5} (b)), the performance of both EWC~\cite{ewc} and online EWC~\cite{rewc} degrade for task $t_{1}$ as do the networks with conventional memory using SGD and ADAM. The FN-synapse based networks on the other hand retain the accuracy of task $t_{1}$ far better in comparison. This advantage in retention comes at the cost of learning $t_2$ marginally poorer than others. This trend of retaining the older memories or tasks far better than other approaches continues in successive tasks. Particularly, if we consider the retention of $t_1$ when the networks are trained on $t_3$ (see Fig. \ref{Fig5} (c)), it can be observed that it is only the FN-synapse based networks that retain $t_1$ while others fall below the $50\%$ threshold. Similar trends can be observed in Fig. \ref{Fig5} (d) and (e). There are a few instances during the five tasks where the EWC variants and SGD with conventional memory marginally outperform or match the FN-synapse in terms of retention. However, if the overall average accuracy of all these approaches are compared (see Fig. \ref{Fig5} (f)), it is clearly evident that both the FN-synapse networks significantly outperform the others. It is also worth noting here that even when a network equipped with FN-synapse is trained using a computationally-inexpensive optimizer such as SGD, it shows remarkably superior performance than highly computationally-expensive approaches such as ADAM with conventional memory and ADAM with EWC variants.  

The only drawback of the FN-synapse based approaches is that its ability to learn the present task slightly degrades with every new task. This phenomenon results from the FN-synapses becoming more rigid and can be seen from Fig. \ref{Fig5} (g) which shows the evolution of plasticity of weights in the output and input layer of the network with successive tasks with respect to $W_{c}$. As mentioned earlier, $W_{c}$ keeps track of the importance of each weight as a function of the number of times it is used. The higher the $W_{c}$ of a particular weight, the less it has been used and therefore, the more plastic it is and sensitive to change. On the other hand, a more rigid and frequently used weight has a lower value of $W_{c}$. If the output layer is considered from Fig. \ref{Fig5} (g), it can be observed that with each successive task the $W_{c}$ of the weights of the network collectively reduces, leading to more consolidation and consequently leaving the network with fewer plastic synapses to learn a new task. In comparison, majority of the weights in the input layer remain relatively more plastic (or less spread out) owing to the redundancies in the network arising from the vanishing gradient problem (see Section \ref{sec12} for more details). 

\section{Methods}\label{met}

\subsection{Weight Update For Differential Synaptic Model}\label{metwup}
\par Consider two dynamical system with state variable $W^+$ and $W^-$ and the rate of change of the state variable governed by a function $J(.)$ . If an external differential time varying input of magnitude $+\frac{1}{2}X(t)$ or $-\frac{1}{2}X(t)$ is applied to the system, then the system dynamics is given by:

\begin{equation}\label{eq3}
\frac{dW^+}{dt}=-J(W^+)+\frac{1}{2}X(t)
\end{equation}
\begin{equation}\label{eq4}
\frac{dW^-}{dt}=-J(W^-)-\frac{1}{2}X(t)
\end{equation}
In this differential architecture, we define the weight parameter $W_{d}$ as $W_{d}=\frac{1}{2}(W^+-W^-)$ which represents the memory and the common-mode parameter $W_{c}$ as $W_{c}=\frac{1}{2}(W^++W^-)$ which represents the usage of the synapse. Applying this definition to \ref{eq3} and \ref{eq4}, we obtain:

\begin{equation}\label{eq5}
\frac{d(W_{c}+W_{d})}{dt}=-J(W_{c}+W_{d})+\frac{1}{2}X(t)
\end{equation}
\begin{equation}\label{eq6}
\frac{d(W_{c}-W_{d})}{dt}=-J(W_{c}-W_{d})-\frac{1}{2}X(t)
\end{equation}
Now, adding and subtracting \ref{eq5} and \ref{eq6}, we get:

\begin{equation}\label{eq7}
\frac{dW_{c}}{dt}=-\left(\frac{J(W_{c}+W_{d})+J(W_{c}-W_{d})}{2}\right)
\end{equation}
\begin{equation}\label{eq8}
\frac{dW_d}{dt}=-\left(\frac{J(W_{c}+W_{d})-J(W_{c}-W_{d})}{2}\right)+X(t)
\end{equation}
Assuming that $W_{c}>>W_{d}$ and applying Taylor series expansion on \ref{eq7} and \ref{eq8}, we get:

\begin{equation}\label{eq9}
\frac{dW_{c}}{dt}=-J\left(W_{c}\right)
\end{equation}
\begin{equation}\label{eq10}
\frac{dW_d}{dt}=-J'\left(W_{c}\right)W_{d}+X(t)
\end{equation}
Substituting the derivative of $W_{c}$ from \ref{eq9} into \ref{eq10}, the rate of change in $W_{d}$ can be formulated as:

\begin{equation}\label{eq11}
\frac{dW_d}{dt}=-\left[\frac{d^{2}W_{c}}{dt^{2}}\left(\frac{dW_{c}}{dt}\right)^{-1}\right]W_{d}+X(t)
\end{equation}
Therefore, the change in weight $\Delta W_{d}$ is directly proportional to the $curvature$ of usage while being inversely proportional to the rate of usage. 
\subsection{Optimal Usage Profile}\label{metopup}
\par We define the decaying term in \ref{eq11} as 

\begin{equation}\label{usage}
r(t)=-\left[\frac{d^{2}W_{c}}{dt^{2}}\left(\frac{dW_{c}}{dt}\right)^{-1}\right]
\end{equation}
Now, comparing the weight update equation in \ref{eq11} to the weight update equation for EWC in the balanced input scenario, the decay term has the following dependency with time for avoiding catastrophic forgetting.

\begin{equation}\label{usa}
r(t)= O\left(\frac{1}{t}\right)  
\end{equation}
Now, the usage of a synapse is always monotonically increasing and since $W_c$ represents the usage, it too needs to monotonic. At the same time $W_c$ also needs to be bounded, therefore $W_c$ has to monotonically decrease with increasing usage while satisfying the relationship in equation \ref{usa}. 
It can be shown that equation~\ref{usa} and~\ref{usage} can be satisfied by any dynamical system of the form 

\begin{equation}\label{genwc}
W_{c} = \frac{1}{f(\log t)}
\end{equation}
where $f(.) \ge 0 $ is any monotonic function. Substituting equation \ref{genwc} in \ref{usage} we obtain the corresponding usage profile as follows

\begin{equation}
r(t)= \frac{1}{t}\left(1+ \frac{2f'(\log t)}{\log t} -\frac{f''(\log t)}{f'(\log t)}\right)  
\end{equation}
where $f'(\log t)$ and $f''(\log t)$ are derivatives of $f(\log t)$ with respect to $\log t$. While several choices of $f(.)$ are possible, the simplest usage profile can be expressed as 

\begin{equation}\label{eqnew}
W_{c} = \frac{\beta}{\log(t)}
\end{equation}
where $\beta$ is any arbitrary constant. The corresponding non-linear function in this model is determined by substituting equation \ref{eqnew} in equation \ref{eq9} to obtain

\begin{equation}\label{eqFN}
J\left(W_{c}\right)=\frac{1}{\beta} W_{c}^2\exp\left(-\frac{\beta}{W_{c}}\right).
\end{equation}
The expression for $J(.)$ in equation~\ref{eqFN} bears similarity with the form of FN quantum-tunneling current and in the next section we discuss the dynamical systems given by equations~\ref{eq3} and~\ref{eq4} can be realized using FN tunneling junctions.

\subsection{Hardware Implementation of Optimal Usage Profile}\label{methardimp}
\par For the differential FN tunneling junctions shown in Fig. \ref{Fig1} (e) and its equivalent circuit shown in the Appendix Fig. \ref{AFig1}, the dynamical systems
model is given by

\begin{equation}\label{eq3_new}
C_T\frac{dW^+}{dt}=-J(W^+)+\frac{C_c}{2}\frac{dx}{dt}
\end{equation}

\begin{equation}\label{eq4_new}
C_T\frac{dW^-}{dt}=-J(W^-)-\frac{C_c}{2}\frac{dx}{dt}
\end{equation}
where $W^+,W^-$ are the tunneling junction potentials, $C_c$ is the input coupling capacitance, $x(t)$ is the input voltage and $C_T = C_c + C_{fg}$ is the total capacitance comprising
of the coupling capacitance and the floating-gate capacitance $C_{fg}$. $J(.)$ are the FN tunneling currents given by

\begin{eqnarray}\label{eqFNcur}
J\left(W^+\right) &=& \left(\frac{k_{1}}{k_{2}}\right) \left(W^+\right)^2\exp\left(-\frac{k_{2}}{W^+}\right) \\
J\left(W^-\right) &=& \left(\frac{k_{1}}{k_{2}}\right) \left(W^-\right)^2\exp\left(-\frac{k_{2}}{W^-}\right)    	
\end{eqnarray}
where $k_{1}$ and $k_{2}$ are device specific and fabrication specific parameters that remain relatively constant under isothermal conditions.
Following the derivations in Section~\ref{metopup} and the expression in equation~\ref{eqnew} leads to a common-mode voltage $W_c$ profile as

\begin{equation}\label{eq2}
W_{c}(t)=\frac{k_{2}}{\log(k_{1}t+k_{0})}    
\end{equation}
where $k_{0}=\exp\left(\frac{k_{2}}{W_{c0}}\right)$ and $W_{c0}$ refers to the initial voltage at the floating-gate.

\subsection{Signal-to-noise Ratio Estimation for Random Pattern Experiment}\label{metsnr}

Following the same procedure in Section~\ref{metopup} the weight update equation for an FN-synapse using equation \ref{eq3_new} and equation \ref{eq4_new} can be expressed as

\begin{equation}\label{eq11_new}
C_T\frac{dW_d}{dt}=-\left[\frac{d^{2}W_c}{dt^{2}}\left(\frac{dW_c}{dt}\right)^{-1}\right]W_{d}+C_c\frac{dx}{dt}
\end{equation}
We designed the floating-gate potential and the input voltage pulses such that the FN-dynamics is only active when there is an memory update. Therefore, the dynamics in equation~\ref{eq11_new} evolve in a discrete manner with respect to the number of modulations. Assuming $C_{T} = C_c$ we formulate a discretized version of the weight update dynamics from equation \ref{eq11_new} in accordance with the floating-gate potential profile of the device expressed in equation \ref{eq2} as follows

\begin{equation}\label{eq12}
\begin{split}
\triangle W_{d}(n)=-k_{1}\left(1+\frac{2}{\log{(k_{1}\triangle tn + k_{0})}}\right)\left(\frac{1}{k_{1}\triangle tn + k_{0}}\right)W_{d}(n-1)\triangle t\\
+\triangle x(n)  
\end{split}
\end{equation}
\begin{equation}\label{eq13}
\begin{split}
& W_{d}(n)=\left[1-\left(1+\frac{2}{\log{(k_{1}\triangle tn + k_{0})}}\right)\left(\frac{1}{n + \frac{k_{0}}{k_{1}\triangle t}}\right)\right]W_{d}(n-1) \\ 
& +(x(n)-x(n-1))    
\end{split}
\end{equation}
where $n$ represents the number of patterns observed and $\Delta t$ is the duration of the input pulse. Let us denote the weight decay term as

\begin{equation}\label{eq14}
\alpha (n) = \left[1-\left(1+\frac{2}{\log{(k_{1}\triangle tn + k_{0})}}\right)\left(\frac{1}{n + \frac{k_{0}}{k_{1}\triangle t}}\right)\right]  
\end{equation}
Thus, we obtain the weight update equation with respect to number of patterns observed as

\begin{equation}\label{eq15}
W_{d}(n)=\alpha(n)W_{d}(n-1) +(x(n)-x(n-1)) 
\end{equation}
When we start from an empty network i.e. $W_{d}(0) = 0$, the memory update can be expressed as a weighted sum over the past input as 

\begin{equation}\label{eq21}
\begin{split}
&W_{d}(n)=\sum_{i=1}^{n-2}\left\{(\alpha(i+1)-1)\left(\prod_{j=i+2}^{n}\alpha(j)\right)x(i)\right\}
+ (\alpha(n)-1)x(n-1)+x(n)
\end{split}
\end{equation}
We define the retrieval signal and the noise associated with it as per the definition in \cite{Ben15}. For a network comprising of N synapses, each weight in the network is indexed as $W_d(a,n)$ where $a=1,...,N$. Similarly, the input applied to the $a^{th}$ synapse after $n$ patterns is $x(a,n)$. Then, the signal strength for the $1^{st}$ update introduced to the empty network tracked after $n$ patterns can be formulated as:

\begin{equation}\label{sigdef}
S(n)=\frac{1}{N}\Bigg\langle\sum_{a=1}^{N}W_{d}(a,n) x(a,1)\Bigg\rangle
\end{equation}
where angle brackets denote averaging over the ensemble of all of the random uncorrelated patterns seen by the network. Substituting equation \ref{eq21} in \ref{sigdef} we obtain

\begin{equation}\label{eq22}
\begin{split}
S(n)=(\alpha(1)-1)\prod_{j=2}^{n}\alpha(j)
\end{split}
\end{equation}
Given that in equation \ref{eq14}, $k_0=\mathcal{O}(10^{19})$ and $k_1=\mathcal{O}(10^{16})$, the term $\left(1+\frac{2}{\ln{(k_{1}\triangle tn + k_{0})}}\right) \approx1$, the signal power simplifies to:

\begin{equation}\label{eq23}
S^2(n)=\frac{1}{\left(n+\gamma\right)^2}\\
\end{equation}
where $\gamma=\frac{k_{0}}{k_{1}\triangle t}$ and depends on the pulse-width $\triangle t$ and the initial condition $k_{0}$. The above equation shows that the signal's strength 
is a function of the system parameter $\gamma$ and decays with the number of memory pattern observed. The corresponding noise power is given by the variance of the retrieval signal expressed in equation~\ref{sigdef}. This can be estimated as the sum of the power of all signals tracked at $n$ except for the retrieval signal corresponding to the first pattern we are tracking and is given by:

\begin{equation}\label{eq24}
\nu^2(n)=\frac{1}{N}\sum_{i=2}^{n}S^{2}(n)
\end{equation}
However, in order to derive a more tractable analytical expression for further analysis we added the retrieval signal as well into the summation which introduces a small error in the estimation (overestimating the noise by the retrieval signal term). This leads us to the following estimation of the noise power:

\begin{equation}\label{neweq24}
\nu^2(n)=\frac{n}{N(n+\gamma)^{2}}
\end{equation}
Based on the value of $n$ in comparison to $\gamma$, we obtain two trends for the noise profile. When $\gamma>>n$,

\begin{equation}\label{eq25}
\nu(n)=\frac{1}{\sqrt{N}}\left(\frac{\sqrt{n}}{\gamma}\right)
\end{equation}
which implies that noise increases with increase in updates initially. On the other hand, when $\gamma<<n$,

\begin{equation}\label{eq26}
\nu(n)=\frac{\sqrt{n}}{\sqrt{N}n}=\frac{1}{\sqrt{N}}\left(\frac{1}{\sqrt{n}}\right)
\end{equation}
which implies that noise falls with increase in updates in the later stages. The signal-to-noise ratio (SNR) of a network of size $N$ can then be obtained as:

\begin{equation}\label{eq27}
SNR(n)=\sqrt{\frac{S^{2}(n)}{\nu^{2}(n)}}=\sqrt{\frac{N}{n}}\\
\end{equation}

\subsection{Programming and Initialization of FN-synapses}\label{metprog}
The potential corresponding to the tunneling nodes $W^+$ and $W^-$ can be accessed through a capacitively coupled node, as shown in Appendix Fig. \ref{AFig1}. This configuration minimizes readout disturbances and the capacitive coupling also acts as a voltage divider so that the readout voltage is within the input dynamic range of the buffer. The configuration also prevents hot-electron injection of charge into the floating gate during readout operation. Details of initialization and programming are discussed in~\cite{Meh20}, so here we describe the methods specific for this work. The tunneling node potential was initialized at a specific region where FN-tunneling only occurs while there is a voltage pulse at the input node and the rest of the time it behaves as a non-volatile memory. This was achieved by first measuring the readout voltage every 1 second for a period of 5 min to ensure that the floating gate was not discharging naturally. During this period the noise floor of the readout voltage was measured to be $\approx 100\mu V$. At this stage, an voltage pulse of magnitude 1 V and duration 1 ms was applied at the input node and the change in readout voltage was measured. If the change was within the noise floor of the readout voltage, the potential of the tunneling nodes were increased by pumping electrons out of the floating gate using the program tunneling pin. This process involves gradually increasing the voltage at the program tunneling pin to 20.5 V (either from external source or from on-chip charge pump). The voltage at the program tunneling pin was held for a period of 30s, after which it was set to 0 V. The process was repeated until substantial change in the readout voltage was observed $(\approx 300\mu V)$ after providing an input pulse. The readout voltage in this region was around 1.8 V. 

\subsection{Hardware and Software Experiments for Random Pattern updates}\label{metsnrhard}
The fabricated prototype contained 128 differential FN tunneling junctions, which corresponds to 64 FN-synapses. However, due to the peripheral circuitry only one tunneling node could be accessed at a time for readout and modification. Now, since the memory pattern is completely random, each synapse can be modified independently without affecting the outcome of the experiment. Therefore, two tunneling nodes were initialized following the method  described in Section ~\ref{metwup}. Input pulses of magnitude 4V and duration 100ms was applied to both the tunneling nodes. The change in the readout voltages were measured, and the region where the update sizes of both the tunneling node would be equal was chosen as the initial zero memory point for the rest of the experiment. The nodes were then modified with a series of 100 {\it potentiation} and {\it depression} pulses of magnitude 4.5v and duration 250 ms and the corresponding weights were recorded. This procedure represented the 100 updates of a single synapse. The tunneling nodes were then reinitialized to the zero memory point and the procedure was repeated with different random series of input pulses representing the modification of other 99 synapse in the network. The first input pulses of each series of modification forms the tracked memory pattern. To modify the value of $\gamma$ the FN-synapses were initialized at a higher tunneling node potential.

\par The software model of the FN-synapse was created by extracting the device parameters $k_1$ and $k_2$ from the hardware prototype.  Modelled using MATLAB, the extracted parameters have been shown to capture the hardware response with an accuracy greater than $99.5\%$ in our previous works \cite{Zho17a,Zho19}. These extracted parameters were fed into a dynamical system which follows the usage profile described in Section~\ref{methardimp} and follow the weight update rule elaborated in Section~\ref{metsnr} to reliably imitate the behaviour of the FN-synapse. The software model network was started with exactly the same initial condition as hardware synapses and subjected to the exact memory patterns used for the hardware experiment for the same number of iterations. The simulation was also extended to 1000 iterations for $\gamma_1$ and the corresponding responses are included in Fig~\ref{Fig3}.

\subsection{Implementing the CFN-synapse}\label{metcfn}
As mentioned earlier in Section \ref{subseccas}, catastrophic forgetting occurs in networks with EWC-like memory consolidation models due to the absence of a balanced pattern retention and forgetting mechanism which is present in the cascade model. Since increase in $retention$ is tantamount to increase in rigidity and $forgetting$ is tantamount to decrease in rigidity, it is necessary to adjust the plasticity/rigidity of the synapse accordingly. From Fig. \ref{Fig2} (d) and (e) notice that $W_c$ decreases monotonically with each new updates which correspondingly makes the synapse only rigid. Therefore, to balance the same, the idea is to keep $W_c$ as steady as possible to keep the synapse plastic as long as possible. This can be achieved by including a `global' parameter $V_{mod}$, associated with all the synapses in the network, which modulates/increases $W_c$ `globally' after every update. Here, the term `globally' refers to all the synapses being modulated equally. Moreover, as illustrated in Fig \ref{Fig1}(f) and the equivalent circuit diagram (see Appendix Fig. \ref{AFig1}), the $V_{mod}$ pin can be used to achieve the same effect in the hardware prototype as well. Note that since the memory is formed as a difference between two tunneling node potential, increasing the potential for all tunneling nodes ensures that differential memory is still preserved.  Now, there can be numerous possible modulation profile to perform this. As a proof of concept, for our experiment (as reported in Fig. \ref{Fig4}) we determined the amount of modulation as half the average of $\Delta W_d$ across all the synapse during the latest update and then used the software model to analyse its comparative performance in MATLAB simulations. 

\subsection{Neural Network Implementation using FN-synapses}\label{metnn}
The MNIST dataset was split into 60,000 training images and 10,000 test images which yielded about 6000 training images and 1000 test images per digit. Each image, originally of 28$\times$28 pixels, was converted to 32x32 pixels through zero-padding. This was followed by standard normalization to zero mean with unit variance.
The code for implementing the non-FN-synapse approaches such as EWC and online EWC were obtained from the repository mentioned in \cite{NIPS18}. To enforce an equitable comparison, the same neural network architecture, in the form a multi-layered perceptron (MLP) with an input layer of 1024 nodes, two hidden layers of 400 nodes each (paired with the ReLU activation function) and a softmax output layer of 2 nodes, has been utilized by every method mentioned in this work. Based on the optimizer in use, a learning rate of 0.001 was chosen for both SGD and ADAM (with additional parameters $\beta_{1}$, $\beta_{2}$ and $\epsilon$ set to 0.9,0.999 and $10^{-8}$ respectively for the latter). Each model was trained with a mini-batch size of 128 for a period of 4 epochs.

Corresponding to every weight/bias in the MLP, an instance of the FN-synapse model was created and initialized to a tunneling region according to the initial $W_{c}$ value. As demonstrated by the measured results in Section \ref{21}, $\Delta W_{d}$ can be modulated linearly and precisely by changing the pulse-width of the {\it potentiation}/{\it depression} pulses. Therefore, each weight update (calculated according to the optimizer in use) is mapped as an input pulse of proportional duration for the FN synapse instance. Then, every instance of the FN-synapse model is updated according to Eq. \ref{eq12} and the $W_{d}$ thus obtained in voltage is scaled back to a unit-less value and within the required range of the network.

\section{Discussion}\label{sec12}
In this paper we reported a differential FN quantum-tunneling based synaptic device that can exhibit near-optimal memory consolidation that has been previously demonstrated using only algorithmic models. The device called FN-synpase, like its algorithmic counterparts, stores the value of the weight and a relative usage of the weight that determines the plasticity of the synapse. Similar to an EWC model, an FN-synapse, `protects' important memory by reducing the plasticity of the synapse according to its usage for a specific task. In this paper we have demonstrated this memory consolidation property of FN-synapse for a benchmark continual learning task. Unlike its algorithmic counterparts like the cascade or EWC models, the FN-Synapse doesn't require any additional computational or storage resources. Memory consolidation in continual learning is achieved by augmenting the loss function using penalty terms that are associated with either Fisher information~\cite{ewc} or the historical trajectory of the parameter over the course of learning~\cite{oewc,rewc}. Thus, the synaptic updates require additional pre-processing of the gradients, which in some cases could be computationally and resource intensive. FN-synapse on the other hand, does not require any pre-processing of gradients and instead can exploit the physics of the device itself for synaptic intelligence and for continual learning. 
For the same benchmark task, we have shown an FN-synapse network shows better multi-task accuracy compared to other continual learning approaches. This leads to the possibility that the intrinsic dynamics of the FN-synapse could provide important clues on how to improve the accuracy of other continual learning models as well. 

Fig. \ref{Fig5} also shows the importance of the learning algorithm in fully exploiting the available network capacity. While the entropy of the FN-synapse weights for the output layer is relatively high, the entropy of the weights of the input layer is still relatively low, implying most of the input layer weights remain unused. This is an artifact of {\it vanishing gradients} in a standard backpropagation based neural network learning. Thus, it is possible that improved backpropagation algorithms~\cite{backprop1,backprop2} might be able to mitigate this artifact and in the process enhance the capacity and the performance of the FN-synapse network. In Appendix Fig. \ref{AFig6} we show that FN-synapse based neural network is able to maintain its performance even when the network size is increased. Thus, it is possible that the network becomes capable of learning more complex tasks due to increase in overall plasticity of the network while ensuring considerably better retention than neural networks with traditional synapses.

\par In addition to being physically realizable, the FN-synapse implementation also allows interpolation between the cascade consolidation and the EWC consolidation models. This is important because it is widely accepted that the EWC model can potentially suffer from blackout catastrophe~\cite{ewc} as the learning network approaches its capacity. During this phase, the network becomes incapable of retrieving any previous memory as well as is unable to learn new ones~\cite{ewc}. Cascade consolidation models and SGD-based continuous learning models avoid this catastrophe by gracefully forgetting old memories. As shown in Fig. \ref{Fig4} (a), an FN-synapse network,
through use of a global modulation factor, is able to interpolate between the two models. In fact the results
in Fig. \ref{Fig4} (b). shows that not only the fraction of patterns/memories retained in an FN-synapse network is higher compared to that of a high-complexity cascade model but the degradation of the network performance past its capability is also more graceful. Even though we have not use the interpolation feature for benchmark experiments, we believe that this attribute is going to provide significant improvements for continuous learning of a large number of tasks. 

\par The interpolation property of FN-synapse could mimic some attributes of metaplasticity observed in biological synapses and dendritic spines~\cite{Mahajan19}. The role of metaplasticity, the second-order plasticity of a synapse which assigns a task-specific importance to every successive task being learned \cite{Laborieux21}, is widely accepted as the fundamental component of neural processes key to memory and learning in the hippocampus \cite{Abraham96,Abraham08}. Since unregulated plasticity leads to runaway effects resulting in previously stored memories to be impaired at saturation of synaptic strength \cite{Brun01}, metaplasticity serves as a regulatory mechanism which dynamically links the history of neuronal activity with the current response \cite{Hulme14}. The FN-synapse mimics the same regulatory mechanism through the decaying term $r(t)$ that takes into account the history of usage or neuronal activity to determine the plasticity of the synapse for future use as well as prevents runaway effects by making the synapses rigid at saturation.

\par Compared to other synaptic devices, FN-synapse is similar to a non-volatile memory (NVM) in the sense that it retains the information (weight) stored in absence of external power supply and in addition, also retains the state of its plasticity. Consequently, this elicits a discussion on its relative merits in comparison with other NVM-based synaptic implementations. Not only does an FN-synapse network minimizes energy requirement by exhibiting continual learning properties without additional computations, it is also quite energy efficient to update the FN-synapse as well. The energy required to update the FN-synapse can be estimated by measuring the energy drawn from the $X(t)$ in Fig.~\ref{Fig1}(e) to charge the coupling capacitors. The magnitude of the coupling capacitor on the fabricated device is $50fF$ and the change in voltage of $4.5V$ which leads to 500fJ per update. Note that the energy required to change the synaptic weight is derived from the FN tunneling current and from the charge initially stored on the floating-gates. Thus, by designing more efficient charge sharing techniques across the coupling capacitors the energy-efficiency of FN-synapse based consolidation can be significantly improved. Furthermore, when implemented on more advanced silicon process nodes, the capacitances could be scaled by an order of magnitude which can further improve the energy-efficiency of FN-synapse based consolidation.

\par Another point of discussion is whether the optimal usage profile presented in this paper can be implemented by other synaptic devices. In recent years, memristors have become the prime candidates for various non-volatile memory designs due to their numerous advantages over more archaic counterparts such as flash \cite{Pal19JEDS}. Of particular interest to us are the memristor-based synapse designs for neuromorphic computing \cite{Mehonic20,Karunaratne20,Tuma16,Fuller19} given the energy-efficient nature of their operation \cite{Pal19MST}. Employing materials such as $Ge_{15}Sb_{85}$ and $Sb$ on $Al_{2}O_{3}$ and $SiO_{2}$ dielectrics in \cite{Sarwat22}, the memtransistive synapse uses a `phase-change' mechanism to switch between two structural states of varying conductivity. The non-volatile and reversible nature of its conductivity has been interpreted as the change of plasticity of the synapse. However, in order to meet the specifications of a targetted neural network, tedious fabrication processes of such memristors are required \cite{Burr17}. Reconfigurable memristors that employ halide perovskite nanocrystals \cite{John22} are more generalized in their application and show capability of prolonged usage of about $5,600$ cycles/updates. In comparison, the FN-synapse has a highly extensive scope of applications as the update size of any network it is employed in can be adjusted appropriately by shaping the input pulse as required. Moreover, in our previous work \cite{adapt}, we have reported that the FN-device can exhibit endurance of about $10^{6}-10^{7}$ cycles without any deterioration. In fact, in order to obtain the memory-lifetime analysis shown in Fig. \ref{Fig3}, a single set of dynamical systems on the hardware prototype (shown in Fig. \ref{Fig1} (d)-(f)) was fed with 100 randomized balanced inputs/updates for 100 cycles to mimic a network of 100 synapses and the was repeated for about 100 Monte-Carlo iterations, which shows an endurance of $10^{6}$ without observing any damage to the oxide layer.

The real limitation of using memristor-based synapses comes from their dynamic range. Generally, a single memristor has two distinct conductive states (corresponding to `0' or `1') which give each device a 1-bit resolution.  When used in a crossbar array, highly-dense designs can reach densities upto $76.5nm^{2}$ per bit as reported by \cite{46nm2perbit} where a 3-D memristor array was constructed using Perovskite quantum wires. The dynamic range or resolution of such designs is determined by the number of memristive devices that can be packed into the smallest feasible physical form factor. If we consider multi-level memristors instead, the resolution per memristor can reach upto 3-5 bits depending on the number of stable distinguishable conductive states \cite{27plus,20level,12level}. In comparison, the dynamic range of the FN-synapse (a single device) is considerably higher as it is determined by the number of electrons stored on the floating-gates which in-turn is determined by the FN-synapse form-factor and the dielectric property of the tunneling barrier. Thus, theoretically the dynamic range and operational-life of the FN-synapse seems to be constrained by the single-electron quantization. However, at low-tunneling regimes the transport of single electrons becomes probabilistic where the probability of tunneling is now modulated by the external signal. In the Appendix Section \ref{asec3} we implemented a stochastic dynamical system emulating single-electron dynamics and driven by random input patterns, similar to the experiment in Section 2.2. Appendix Fig. \ref{AFig3} shows that the SNR still follows the optimal power law curve and the FN-synapse network continues to learn new experiences even if the synaptic updates are based on discrete single-electron transport. 

A more pragmatic challenge in using the FN-synapse will be the ability to discriminate between changes in floating-gate voltage which would be in the order of 100nV per electron. A more realistic scenario would be to measure the change in voltage after 1000 electron tunneling events which would imply measuring 100 $\mu$V changes. Although this will reduce the resolution of the stored weights/updates to 14 bits, recent studies have shown that neural networks with training precisions as low as 8 bits~\cite{8bittrain} and networks with inference precisions as low as 2-4 bits~\cite{2bitinf,4bitinf} are often capable of exhibiting remarkably good learning abilities. Therefore, a precision of 14 bits can be expected to be more than good enough for training large neural networks. 









\section{Appendix}

\subsection{Equivalent Circuit Model of FN-Synapse}\label{asec1}

 \begin{figure}[h]
    \centering
    \includegraphics[width=12cm]{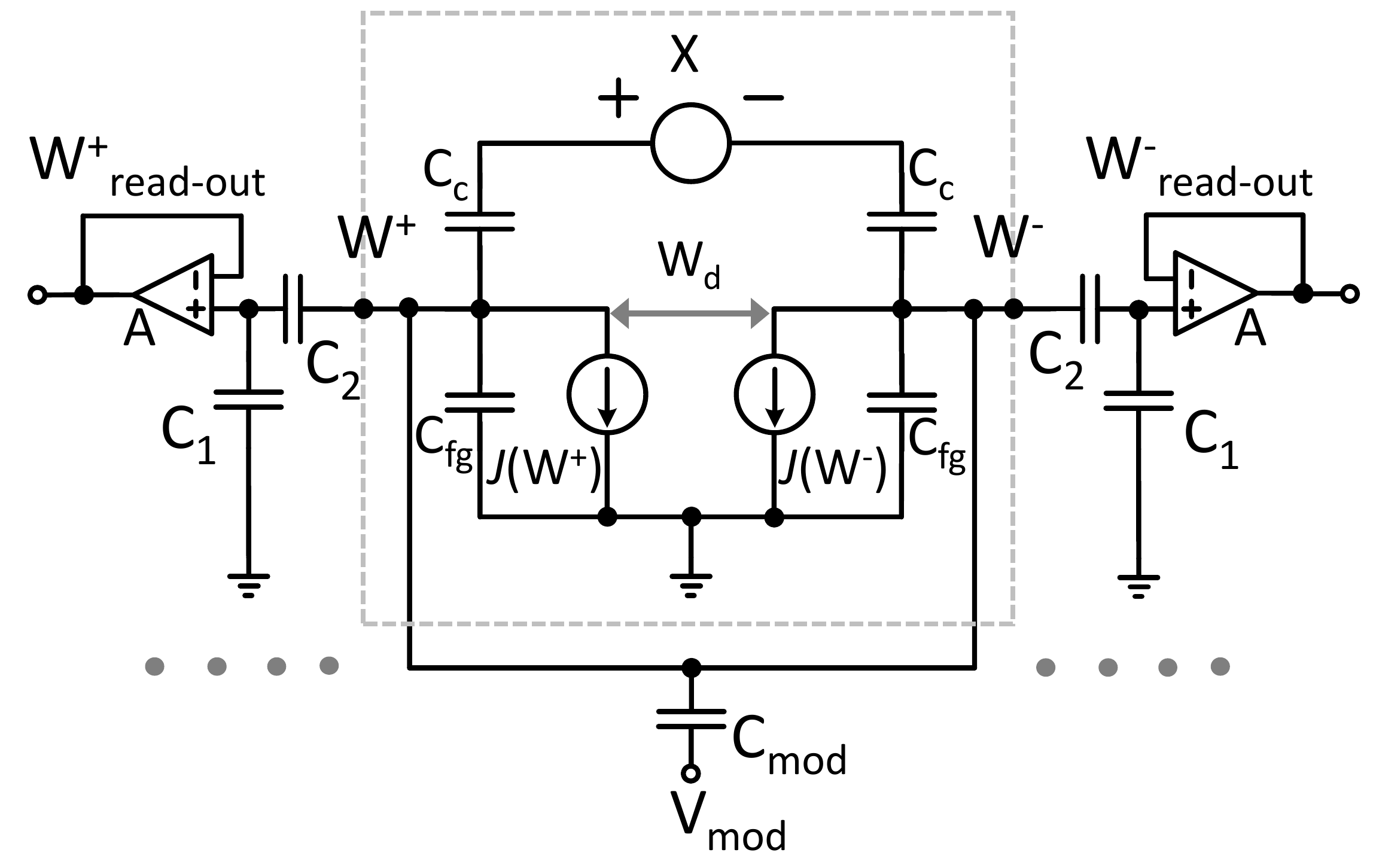}
    \caption{Equivalent circuit model of an FN-synapse.}\label{AFig1}
 \end{figure}
The equivalent circuit model of a single FN-synapse is shown in Appendix Fig. \ref{AFig1}. The synaptic weight $W_d$ is stored as a difference between the voltages ($W^+$ and $W^-$) on the floating-gates. The FN tunneling current is modeled using voltage dependent current sources $J(W^+)$, $J(W^-)$ that discharge the floating-gate capacitances $C_{fg}$. Both $W_d$ and the common-mode voltage $W_c$ are estimated by measuring $W^+$ and $W^-$ using
a capacitive divider formed by $C_1$ and $C_2$ and respective source-followers $A$. This configuration has been previously demonstrated to avoid read-disturbances when measuring the floating-gate voltages~\cite{Meh20,Zho17a}. External input $X$ is differentially coupled to the FN-synapse through the capacitances $C_c$ and $C_{mod}$ is used to couple the signal $V_{mod}$ common to all synapses. $V_{mod}$ is used to adjust the plasticity of the entire synaptic array. The initial charge on the floating-gates are programmed using a combination of FN quantum-tunneling and hot-electron injections, details of which can be found in~\cite{Meh20}.  

\subsection{Modeling Results}\label{asec2}

\begin{figure}[h]
    \centering
    \includegraphics[width=12cm]{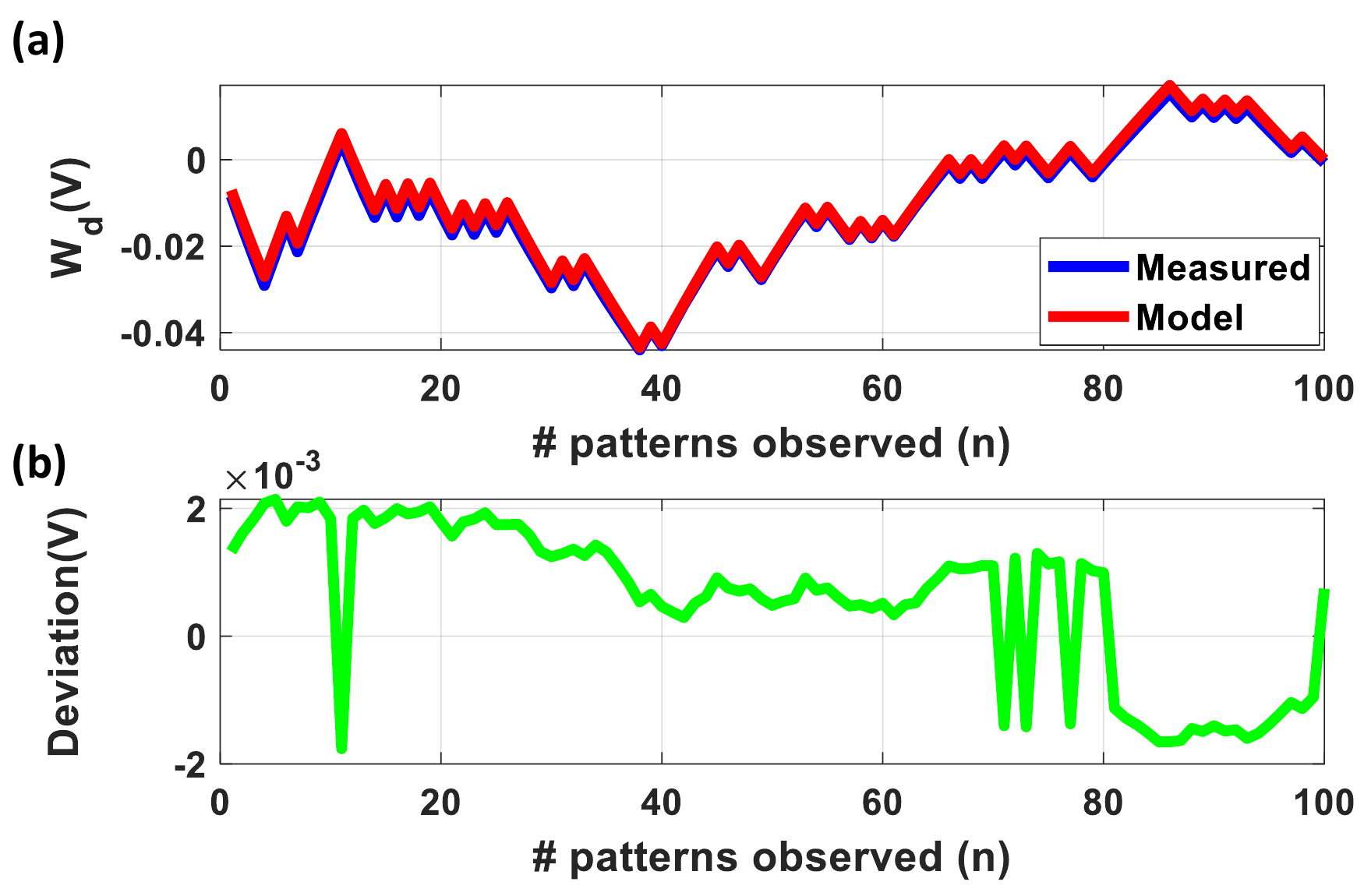}
    \caption{(a) Comparison of weight ($W_{d}$) stored in the FN-synapse and its software model of equivalent plasticity and initial conditions when exposed to the same input pattern and (b) the corresponding deviation.}\label{AFig8}
    \end{figure} 
\subsubsection{Behavioral Model of the FN-Synapse}\label{38}
\begin{figure}[h]
    \centering
    \includegraphics[width=12cm]{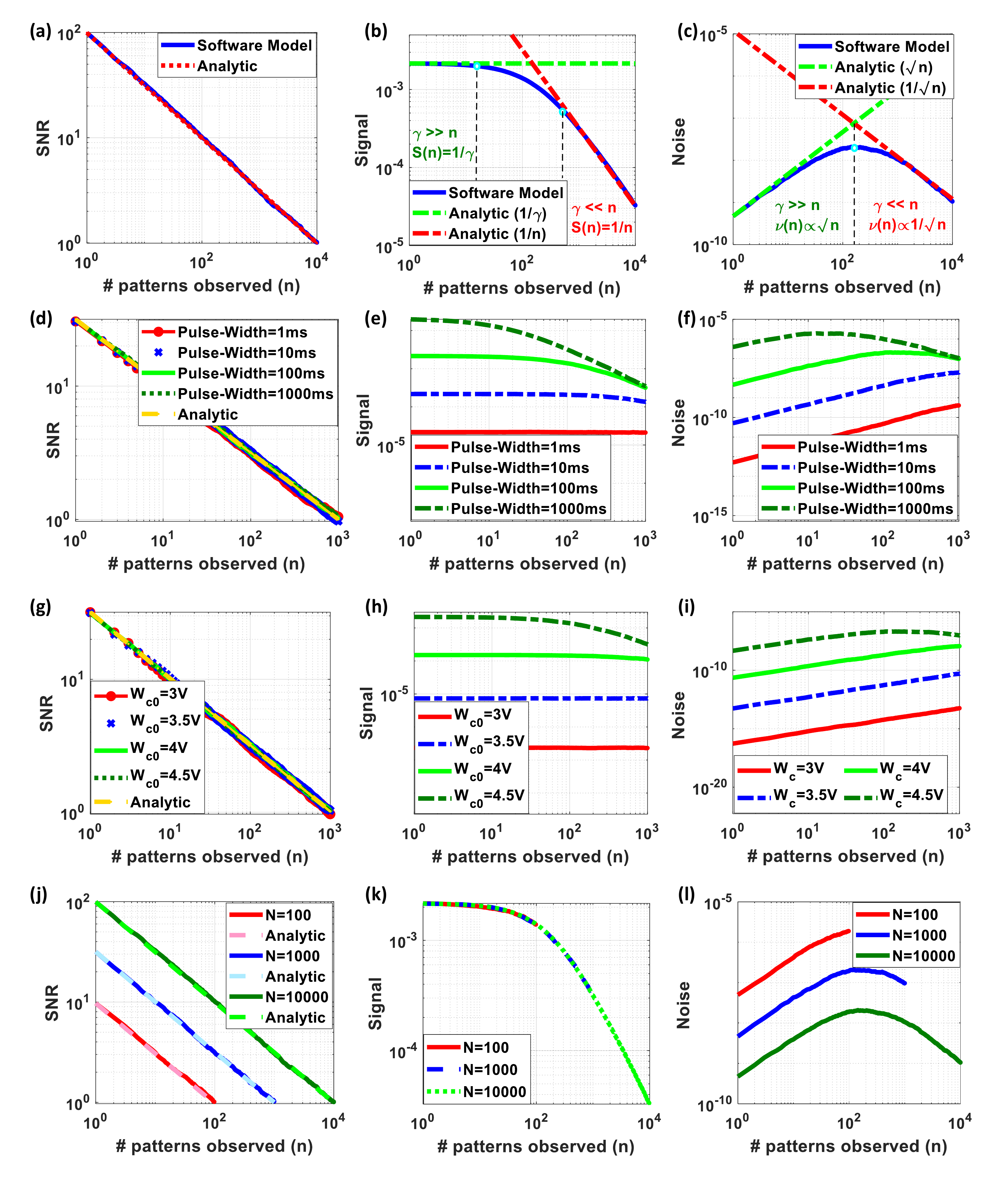}
    \caption{Comparison between the behavioral model and the analytical model of the FN-synapse in terms of (a) SNR, (b) signal and (c) noise. The effect on the SNR, signal and noise of the software model when (d)-(f) the pulse-width of the input pulse is varied and when (g)-(i) the magnitude of the input pulse is varied. (j)-(l) The impact of change in network size on SNR, signal and noise .}\label{AFig2}
\end{figure}
The fabricated prototype of the FN-synapse array comprises of 64 FN-synaptic elements. Thus, for large-scale memory consolidation experiments and for large-scale continual learning experiments, we require a behavioral model that can accurately capture the response of each FN-synapse in the array. In our previous works ~\cite{Zho19,Zho17a} we have validated that equation \ref{eqnew} in the main manuscript can accurately (accuracy greater than $99\%$) model the dynamic response of a single FN tunneling junction and a corresponding integrator. For this work we instantiated two tunneling junctions corresponding to the floating-gates $W^+$ and $W^-$ and the model parameters $k_0$, $k_1$ and $k_2$ were estimated using measured results. A non-linear regression was specifically used to estimate $k_1$ and $k_2$~\cite{Meh20,Zho17a}, whereas $k_0$ was determined from the voltage to which each of the floating-gates were initialized. To validate the behavioral model of the FN-synapse, we carried out a set of experiments and compared the outputs against the analytical results shown in the Methods Section. Appendix Fig \ref{AFig8} and \ref{AFig2} summarizes all the results obtained from the behavioral model.
\par In the first experiment, we measured the weight evolution of an FN-synapse using the fabricated prototype for a series of potentiation/depression pulses. The same input was provided to the software model and the weight evolution was simulated. Appendix Fig~\ref{AFig8} (a) shows that the stored weight of the software model accurately matches with that of the hardware FN-synapse with a small deviation as shown in Fig~\ref{AFig8} (b). This verifies that both hardware FN-synapse and software model behaves similarly when subjected to same stimuli. Next, we ran a Monte-Carlo simulation where we updated a network of $N=10000$ FN-synapses with random binary pattern. Each tunneling junction of FN-synapses were initialized at $W_{c0}=4.5v$. The updates were provided as a differential input voltage pulses of magnitude $4V$ and duration $\Delta t = 100mS$ to each synapses. The experiment was repeated for $1000$ Monte-Carlo simulations. Appendix Fig \ref{AFig2} (a), (b), and (c) shows the SNR, memory retrieval signal $S(n)$ and the noise $\nu(n)$ respectively obtained from the software model of FN-synapse network. In Appendix Fig \ref{AFig2} (a) we observe that the SNR from the software model matches accurately with the analytical expression. Both $S(n)$ and $\nu(n)$ described in equation \ref{eq1} in the main manuscript have two different regimes depending on the value of $\gamma$. When $n<<\gamma$, $S(n)$ is approximately constant and $\nu(n)$ increases at a rate of $\sqrt{n}$. On the other hand, when $n>>\gamma$, $S(n)$ and $\nu(n)$ falls off at a rate of $\frac{1}{n}$ and $\frac{1}{\sqrt{n}}$ respectively. Appendix Fig \ref{AFig2} (b) and (c) shows that the response from the software model follows theses trends and captures both the regimes accurately. 
\par In the next set of numerical experiments, we verified whether the FN-synapse network shows similar trends as the analytic expression in response to changing the value of $\gamma$ in equation \ref{eq1} in the main manuscript. Note that the parameter $\gamma$ is defined as
\begin{equation}
    \gamma = \frac{k_0}{k_1\Delta t}
\end{equation}
where $k_0=exp(\frac{k_2}{W_{c0}})$. Therefore, $\gamma$ for the same set of FN-synapses increases when $\Delta t$ or $W_{c0}$ decreases and vice versa. According to equation \ref{eq15} and \ref{eq21}, the value of $n$ at which the regimes in these responses changes also shifts. Moreover, the initial values for both $S(n)$ and $\nu(n)$ depends on the value of $\gamma$ while SNR is agnostic to changes in $\gamma$. Appendix Fig \ref{AFig2} (d)-(i) show the FN-synapse responses in relation to changing the pulse width and the initialization condition for a network size of $N=1000$. From the figures we can observe that the software model is in very good agreement with the analytic expressions. Finally, we verify the behavioral model in relation to change in the size $N$ of the FN-synapse network. From the analytic expressions in equation \ref{eq1} in the main manuscript, $SNR\propto \sqrt{N}$ and $\nu(n)\propto \frac{1}{\sqrt{N}}$ while $S(n)$ remains constant with respect to $N$. Appendix Fig \ref{AFig2} (j)-(l) shows that the FN-synapse network exhibits these attributes accurately. Note that the regime switching point in $S(n)$ and $\nu(n)$ remains constant, since $\gamma$ does not depend on the size of the network.

\subsubsection{Probabilistic FN-Synapse Model }\label{39}
    \begin{figure}[h]
    \centering
    \includegraphics[width=12cm]{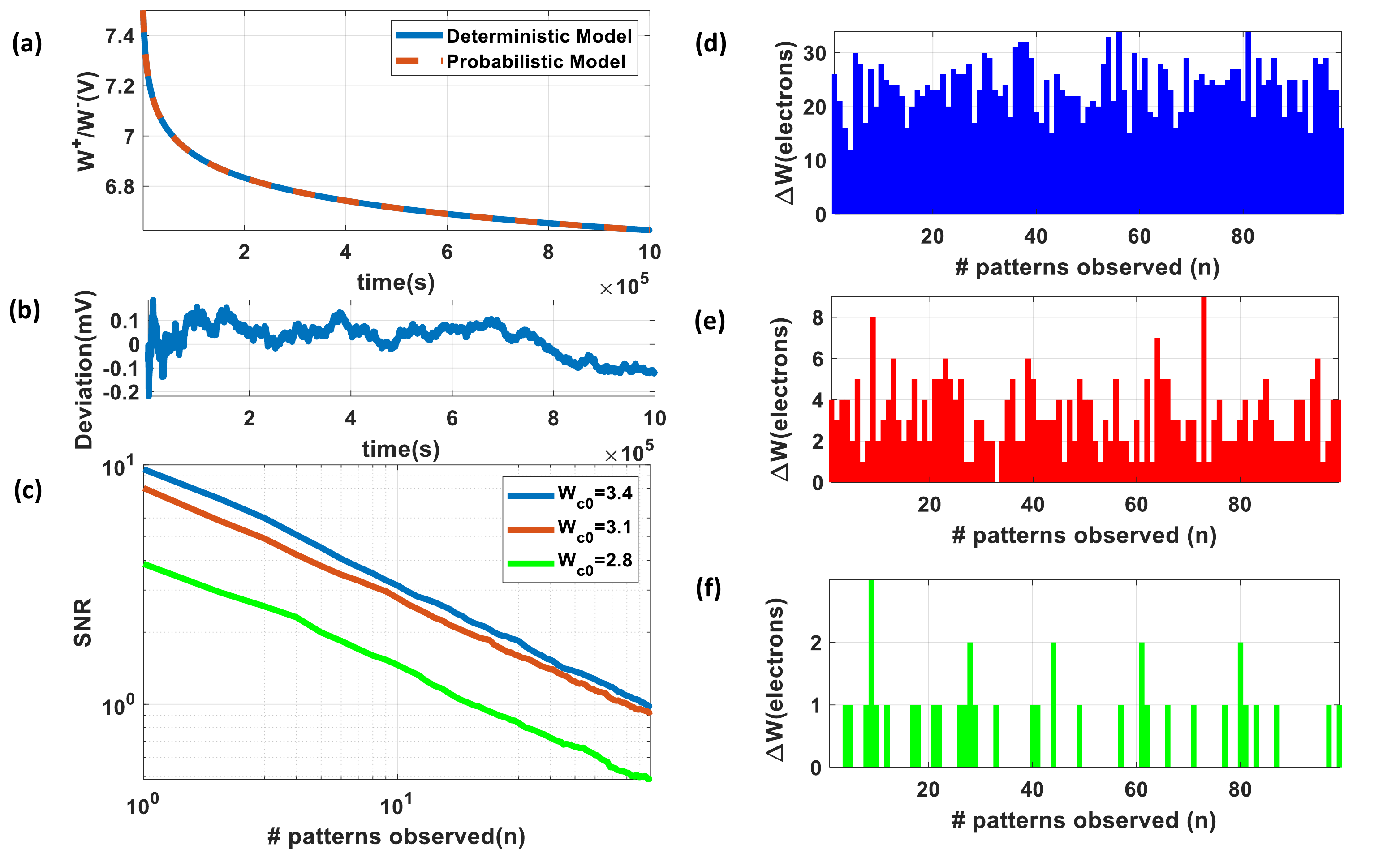}
    \caption{(a) Comparison between the output of the probabilistic FN-synapse model and the deterministic behavioral model and the (b) corresponding deviation. (c) The SNR of the network for different tunneling regions for $W_{c0}$ = 3.4V, 3.1V and 2.8V and (d)-(f) their corresponding update size in terms of no. of electrons per update.}\label{AFig3}
    \end{figure}
    The update process for FN-synapse involves tunneling of electron through a triangular FN quantum-tunneling barrier. The tunneling current density is dependent on the barrier profile which in turn is a function of the floating-gate potential. When $W^+,W^-$ is around 7 V the synaptic update $\Delta W_d$ due to an external pulse can be found out using the continuous and deterministic form of the FN-synapse model (as described in the previous section). Since the number of electrons tunneling across the barrier is relatively large ($\gg 1$), the method is adequate for determining $\Delta W_d$. However, once $W^+,W^-$ is around 6 V, each updates occurs due to the transport of a few electrons tunneling across the barrier and in the limit only one electron tunneling across. In this regime, the continuous behavioral model is no longer valid. Therefore, in this region the FN-synapse switches to a probabilistic model. We can assume that each electron tunneling event follows a Poisson process where the number of electrons $e^+(n),e^-(n)$ tunneling across the two junctions during the $n^{th}$ input pulse is estimated by sampling from a Poisson distribution with rate parameters $\lambda^+,\lambda^-$ given by
    \begin{eqnarray}\label{poi}
    	\lambda^+(n) &=& \frac{AJ(W^+(n))}{q} \\
    	\lambda^-(n) &=& \frac{AJ(W^-(n))}{q}.
    \end{eqnarray}
    $q$ is the charge of an electron, $A$ is the cross-sectional area of the tunneling junction. 
    Using the sampled values of $e^+(n),e^-(n)$, the corresponding discrete-time stochastic equation governing the dynamics of the tunneling node potentials $W^+(n),W^-(n)$ is given by
    \begin{eqnarray}
        W^+(n) &=& W^+(n-1)-\frac{qe^+(n)}{C_{T}} \\
        W^-(n) &=& W^-(n-1)-\frac{qe^-(n)}{C_{T}} 
    \end{eqnarray}
    where $C_{T}$ is the equivalent capacitance of the tunneling node.
    \par We have verified the validity/accuracy of the probabilistic model against the continuous-time deterministic model in high tunneling rate regimes. Appendix Fig. \ref{AFig3} (a) shows that the output of the probabilistic model matches closely to the deterministic model and the deviation which arises due to the random nature of the probabilistic updates (shown in Appendix Fig. \ref{AFig3} (b)) is within $200\mu v$. Using the probabilistic model we performed the memory retention and network capacity experiments (as discussed in the main manuscript) by initializing the tunneling nodes at a low potential. In this regime, each updates to the FN synapse results from tunneling of a few electrons. Appendix Fig. \ref{AFig3} (c) and (d) shows that even when each update sizes are on the order of tens of electrons, the network capacity and memory retention time remains unaffected. However, as the update sizes go below ten electrons per modification (shown in Appendix Fig. \ref{AFig3} (e)), the SNR curve starts to shift downwards and the network capacity along with memory retention time decreases. The tunneling node potential can be pushed further down to a region where the synapses might not even register modifications at times and other times update sizes drop down to single electron per modification (see Appendix Fig. \ref{AFig3} (f)). In this regime, the SNR curve shifts down further, the SNR decay still obeys the power-law curve.
    
    \subsection{Plasticity and Consolidation}\label{asec3}
    \begin{figure}[h]
    \centering
    \includegraphics[width=12cm]{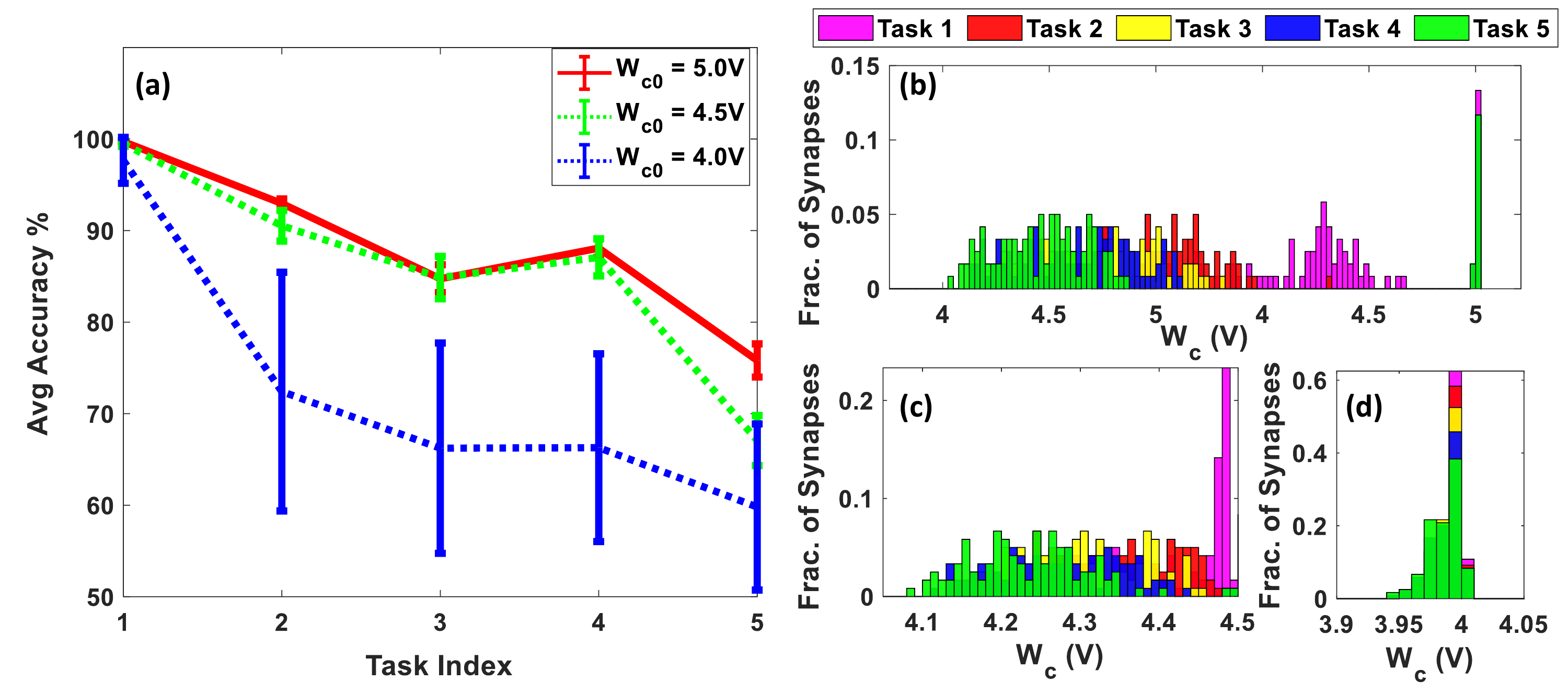}
    \caption{Effect of initial plasticity ($W_{c0}$) of FN-synapse on (a) overall average accuracy of the split-MNIST incremental domain learning tasks as a result of the degree of change in plasticity of their corresponding weights for (b) $W_{c0}$ = 5.0V, (c) $W_{c0}$ = 4.5V and (d) $W_{c0}$ = 4.0V.}\label{AFig4}
    \end{figure}
    The ability of a network to learn new tasks is contingent on the availability of adequate range of plasticity of the synapses so that the weights learned from previous tasks can adapt sufficiently to reflect the requirements for the new tasks. Traditional volatile memories have practically infinite range of plasticity and can therefore change the weights stored to any extent that is required. However, this feature might not be beneficial for continual learning where the network needs to learn new tasks without forgetting the previous ones. This {\it rigidity-plasticity} dilemma is a the core underpinning of memory consolidation where more frequently used synapses become more rigid in comparison to the less frequently used synapses. Thus, a balance between the range of plasticity required to learn successive tasks and the consolidation of the weights learned in the process is key to continual learning. In the case of FN-synapse based neural networks, the range of plasticity is determined by the initial tunneling region of the device. A high tunneling region, denoted by a larger value of $W_{c0}$, ensures that the synapses are plastic enough to learn several successive tasks and slowly become rigid over time. This is seen in the case of $W_{c0}$ = 5V and $W_{c0}$ = 4.5V, which exhibit significantly better overall average accuracy over five tasks as shown in Appendix Fig. \ref{AFig4} (a) as the weights stored in their synapses (shown in Appendix Fig. \ref{AFig4} (b) and \ref{AFig4} (c) respectively) slowly spread from a highly plastic to a rigid region over the course of the five tasks. In contrast, a relatively low initial tunneling region, such as in the case of $W_{c0}$ = 4V, does not learn new tasks as well as the previous couple of cases as shown in Appendix Fig. \ref{AFig4} (a) since in this case the weights stored in the synapse are already relatively rigid at the point of initiation and barely undergo any change as illustrated in Appendix Fig. \ref{AFig4} (d). Therefore by choosing the initial plasticity level appropriately we can achieve an optimal balance between the range of plasticity and consolidation suitable for continual learning.
 
\subsection{Neural Network Architecture}\label{asec5}
    \begin{figure}[h]
    \centering
    \includegraphics[width=12cm]{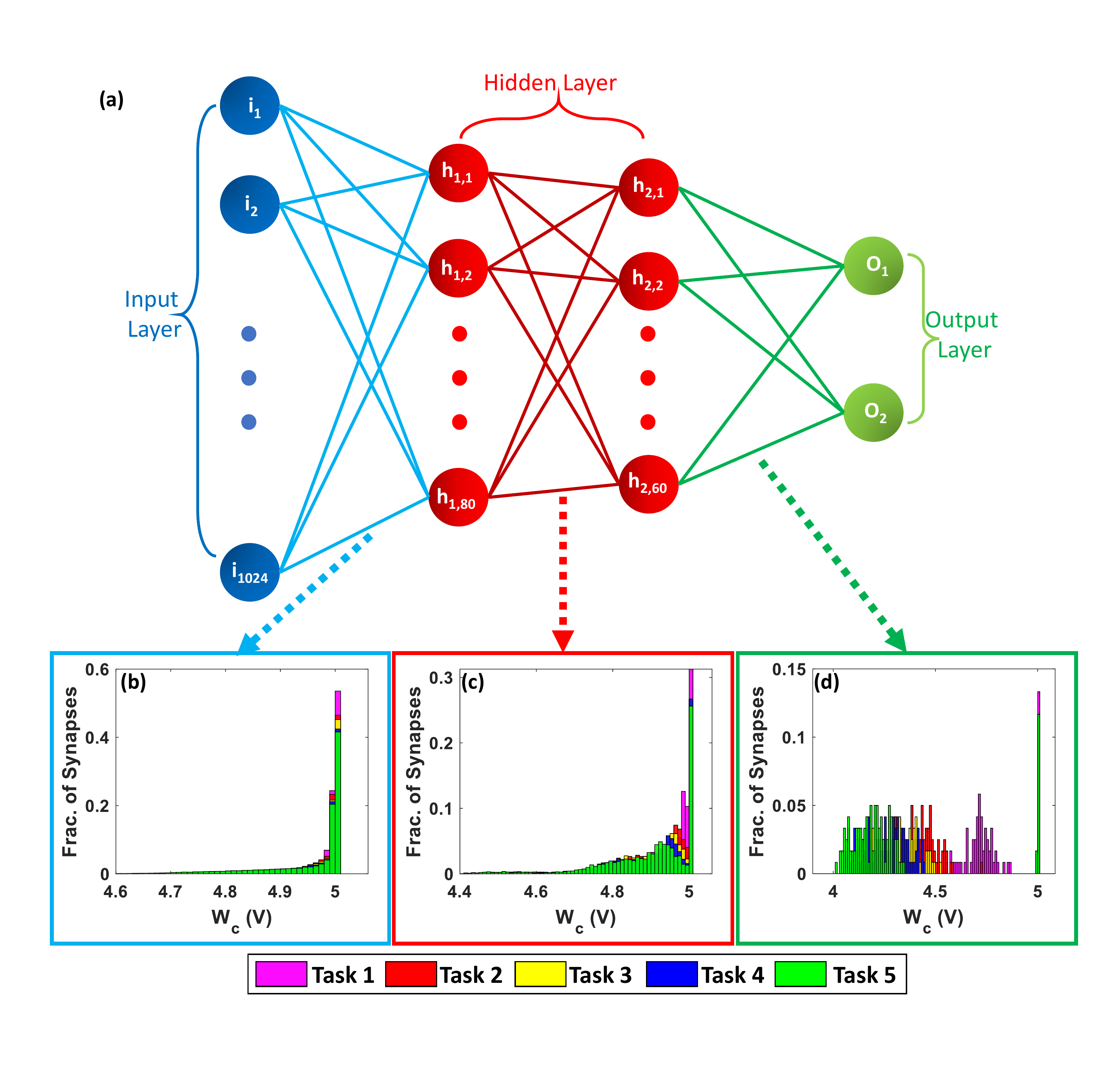}
    \caption{(a) The architecture of the neural network used in the report and the evolution of corresponding weights in between (b) layer 1 and 2, (c) layer 2 and 3, and (d) layer 3 and 4 over five successive tasks.}\label{AFig5}
    \end{figure}
    \begin{figure}[h]
    \centering
    \includegraphics[width=12cm]{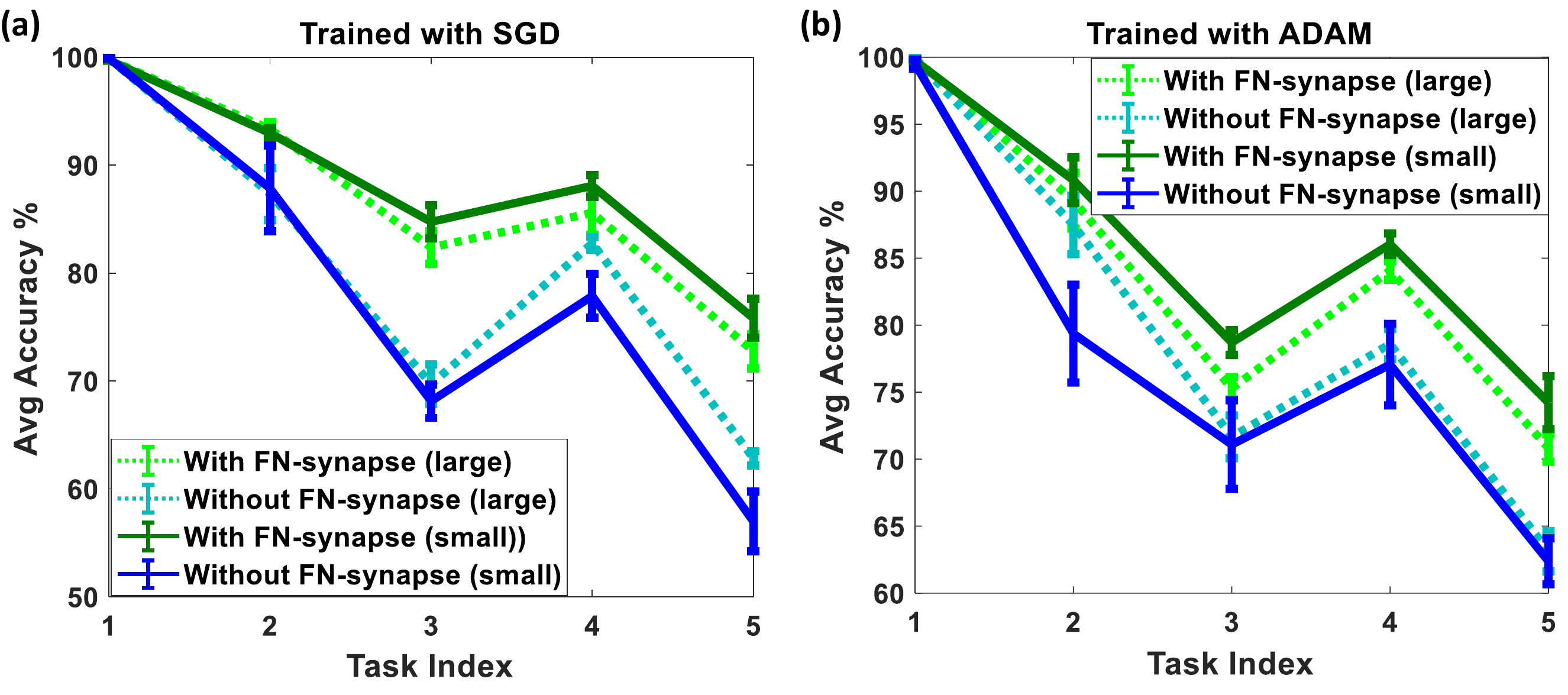}
    \caption{Effect of network size on overall average accuracy when trained with (a) SGD and (b) ADAM.}\label{AFig6}
    \end{figure} 
    The architecture of the 4-layer fully-connected MLP is shown in Appendix Fig. \ref{AFig5} (a). Comprising an input layer of 1024 neurons corresponding to images of 32x32 pixels, two hidden layers of 80 and 60 neurons each and an output layer of 2 neurons that differentiates between (0,1) in $t_{1}$, (2,3) in $t_{2}$, (4,5) in $t_{3}$, (6,7) in $t_{4}$ and (8,9) in $t_{5}$ the network was constructed in MATLAB and trained with SGD and ADAM with learning rate of 0.001 for 4 epochs with a mini-batch size of 128. For comparisons with EWC and Online EWC, the network was replicated in python and trained with exactly the same parameters.
    
        The evolution of the plasticity/usage of weights of the different layers of the FN-synapse based neural network are shown in Appendix Fig. \ref{AFig5} (b)-(d). Given the relatively large number of weights between layer 1-2 and layer 2-3, the amount of change in plasticity that they undergo (as shown in Appendix Fig \ref{AFig5} (b) and \ref{AFig5}(c) respectively) is much lesser in comparison with those between layer 3-4 (as shown in Appendix Fig \ref{AFig5} (d)) as the presence of much fewer weights ensures that they are modified considerably frequently due to lack of any redundancy.
    
    Fig. \ref{Fig5} of the main manuscript already depicts the advantages of the FN-synapse based neural networks using either SGD or ADAM as the optimizer when employed within the aforementioned architecture. In addition, if the size of the neural network is increased by increasing the number of neurons in the hidden layers from 80/60 in layer 2/3 to 400/400, it can be observed from Appendix Fig. \ref{AFig6} (a)-(b) that the average overall accuracy of the FN-synapse based network still outperforms the ones without it as the memory element. Interestingly, the accuracy of the larger network with FN-synapse is slightly lower than that of the smaller network with FN-synapse for task 3 and beyond. This dip is actually an indication of higher plasticity, and therefore slower consolidation, of the larger network due to presence of many more synapses which are still highly plastic after several tasks, which makes FN-synapse based large neural networks equipped with the capability of learning more complicated tasks than split-MNIST and yet exhibit far better consolidation than conventional memory.

\subsection{Effects of Mismatch}\label{asec6} 
    \begin{figure}[h]
    \centering
    \includegraphics[width=12cm]{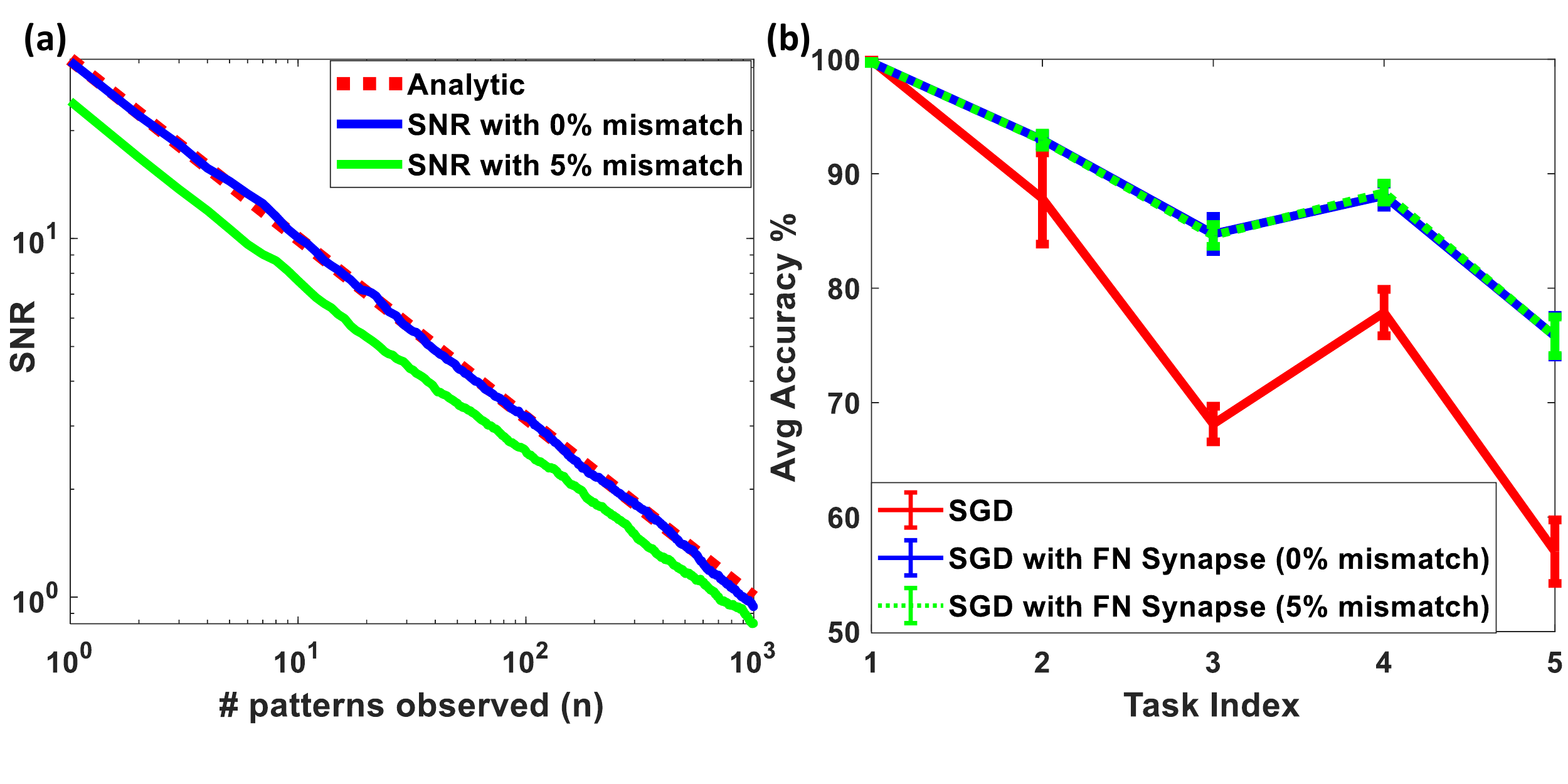}
    \caption{Effect of mismatch in device characteristics across FN synapses on (a) memory retention and (b) learning ability on the split-MNIST based incremental domain learning tasks.}\label{AFig7}
    \end{figure} 
    The FN-synapse comprises of two differential FN tunneling junctions and the operation of the synapse assumes that the junctions are well matched. This will ensure that the weights stored in the synapse are equally plastic/rigid, when increasing or decreasing the magnitude of the weight. A key requirement is that the tunneling rates of the two junctions corresponding to $W^+$ and $W^-$ are synchronized with each other. Previously, we have shown in \cite{adapt,Meh20} that two such FN-dynamical systems can be synchronized to a very high degree of accuracy even in the presence of temperature variations or device mismatch. This is particularly evident from Fig. \ref{Fig3} (d) of the main manuscript, which shows the evolution of weights in a hardware experiment when presented with a 100 randomly balanced input pulse, wherein the uniform change in weights in both the positive and negative directions. A lack of synchronization across the two differential junctions would have resulted in a bias along one of the update directions, which is not observed in the measured results.  
  
    On the other hand, mismatch in device characteristics across one or more FN synapses, specifically the parameters $k_{1}$ and $k_{2}$, must be taken into consideration. This is because a neural network could comprise of billions of synapses and mismatch in synaptic behavior could pose a problem. Appendix Fig. \ref{AFig7} (a) shows the effect of a 5$\%$ mismatch in device characteristics across synapses on the SNR of an FN-synapse network comprising of 10,000 synapses. In this experiment, the network was subjected to 10,000 randomized balanced updates, similar to the previous consolidation experiments. It can be observed that the network with mismatch shows a small degradation in SNR or memory retention compared to the one without any mismatch. However, the SNR still follows the  power-law curve. On the contrary a mismatch of 5$\%$ does not lead to any deterioration whatsoever of the average overall accuracy of the network when trained with SGD over the split-MNIST dataset with the incremental domain learning tasks as depicted in Appendix Fig \ref{AFig7} (b). This shows the robustness of the FN-synapse based network and the ability of learning to compensate for device mismatch.

\bibliography{sn-bibliography}

\section{Acknowledgements}
This work was supported in part by the National Science Foundation grant (ECCS: 1935073).
\section{Author Contributions}
S.C and M.R came up with the concept of FN-synapse. M.R., S.B. and S.C. designed the hardware and simulation experiments, M.R. designed the 64 element FN-synapse chipset; M.R. and S.B. conducted the simulation and hardware experiments. S.C. provided supervision on all tasks. All authors contributed towards writing and proof-reading the manuscript.
\section{Competing Interests}
The authors declare no competing interests.
\end{document}